\documentclass[preprint,12pt]{elsarticle}
\usepackage{amssymb}
\usepackage{float}
\usepackage{amsmath}
\usepackage{caption}
\usepackage{subcaption}
\usepackage[colorlinks,linkcolor=red,anchorcolor=blue,citecolor=green]{hyperref}
\usepackage{geometry}
\usepackage{mathrsfs}

\journal{Elsevier}

\begin{document}

\begin{frontmatter}



\title{Gas-Kinetic Scheme for Partially Ionized Plasma in Hydrodynamic Regime}


\author[a]{Zhigang PU}
\ead{zpuac@connect.ust.hk}
\author[b]{Chang LIU}
\ead{liuchang@iapcm.ac.cn}
\author[a,c]{Kun XU\corref{cor1}}
\ead{makxu@ust.hk}
\cortext[cor1]{Cooresponding author}

\affiliation[a]{organization={Department of Mathematics, Hong Kong University of Science and Technology},
            addressline={Clear Water Bay, Kowloon},
            city={Hong Kong},
            country={China}}
\affiliation[b]{organization={Institute of Applied Physics and Computational Mathematics},
            city={Beijing},
            country={China}}
\affiliation[c]{organization={Shenzhen Research Institute, Hong Kong University of Science and Technology},
            city={Shenzhen},
            country={China}}

\begin{abstract}
Most plasmas are only partially ionized. To better understand the dynamics of these plasmas, the behaviors of a mixture of neutral species and plasma in ideal magnetohydrodynamic states are investigated. The current approach is about the construction of coupled kinetic models for the neutral gas, electron, and proton, and the development of the corresponding gas-kinetic scheme (GKS) for the solution in the continuum flow regime.
The scheme is validated in the 1D Riemann problem for an enlarged system with the interaction from the Euler waves of the neutral gas and magnetohydrodynamic ones of the plasma. Additionally, the Orszag-Tang vortex problem across different ionized states is tested to examine the influence of neutrals on the MHD wave evolution. These tests demonstrate that the proposed scheme can capture the fundamental features of ideal partially ionized plasma, and a transition in the wave structure from the ideal MHD solution of the fully ionized plasma to the Euler solution of the neutral gas is obtained.
\end{abstract}




\end{frontmatter}


\section{Introduction}
\label{introduction}

Partially ionized plasmas(PIP) are a ubiquitous form of matter found throughout the universe, consisting of neutral particles and charged species of ions and electrons. PIP plays significant roles in the fields of astrophysics, low-temperature plasma, aerospace engineering, and so on. In astrophysics, PIP exists in a variety of stellar environments, including the solar chromosphere, Earth's ionosphere, and molecular clouds\cite{ballester2018,popescu_braileanu_two-fluid_2019,khomenko_three-dimensional_2018,soler_theory_2022,balsara_wave_1996,kuzma_numerical_2020,ballai_linear_2019,ballester_nonlinear_2020}. The degree of ionization in these plasmas varies from nearly none in cold regions to almost complete in hot regions. PIP exhibits distinct physical behavior from fully ionized plasma in these cases, such as the decay of the Alfven wave\cite{balsara_wave_1996}, cut-off mode\cite{ballester2018,wojcik_acoustic_2018,alharbi_waves_2022}, Biermann battery effects\cite{martinez-gomez_simulations_2021}, and heating due to ion-neutral friction. With respect to low-temperature plasma, PIP prevails in engineering applications such as microelectronic fabrication\cite{song_control_nodate,quang-zhi_picmcc_nodate,zhang_numerical_2021}, material processing\cite{capitelli_nonequilibrium_1990}, and medical treatment\cite{park_electron_2019,laroussi_cold_2020}. Notably, electrons in these cold plasmas are much hotter than heavy particles like ions and neutral atoms, exhibiting considerable kinetic effects.
Aerospace applications, such as plasma-based flow control, high-speed flows interplanetary reentry, and ion thrusters  \cite{shang_computational_nodate,kaganovich_kinetic_nodate,lefevre_magnetohydrodynamic_2022,xie_experimental_2023,otsu_reentry_2004,katsurayama_particle_2011}, require deep understanding of PIP.

This paper targets an ideal case in the continuum flow regime. The dynamic of neutrals is governed by the Euler equations and the charged species follow the ideal Magnetohydrodynamics(MHD) equations. The coupled system with interaction can be written as,
\begin{equation}
\begin{aligned}
& \partial_t \rho_n+\nabla_{\boldsymbol{x}} \cdot\left(\rho_n \boldsymbol{U}_n\right)=0, \\
& \partial_t\left(\rho_n \boldsymbol{U}_n\right)+\nabla_{\boldsymbol{x}} \cdot\left(\rho_n \boldsymbol{U}_n \boldsymbol{U}_n+p_n \mathbb{I}\right)=S_n, \\
& \partial_t \mathscr{E}_n+\nabla_{\boldsymbol{x}} \cdot\left(\left(\mathscr{E}_n+p_n\right) \boldsymbol{U}_n\right)=Q_n,\\
& \partial_t \rho_c+\nabla_{\boldsymbol{x}} \cdot(\rho_c \boldsymbol{U}_c)=0, \\
& \partial_t(\rho_c \boldsymbol{U}_c)+\nabla_{\boldsymbol{x}} \cdot(\rho_c \boldsymbol{U}_c \boldsymbol{U}_c+p_c I)= (\nabla\times\boldsymbol{B}) \times \boldsymbol{B} + S_c, \\
& \partial_t \mathscr{E}_c+\nabla_{\boldsymbol{x}} \cdot((\mathscr{E}_c+p_c) \boldsymbol{U}_c)=\boldsymbol{J}\cdot\boldsymbol{E} +  Q_c,\\
& \partial_t \boldsymbol{B}+\nabla_{\boldsymbol{x}} \times(\boldsymbol{U}_c \times \boldsymbol{B})=0,
\label{eq:Euler-MHD}
\end{aligned}
\end{equation}
where, subscript $n$ and $c$ stands for neutrals and charged species. $S$ and $Q$ represent momentum and energy exchange between neutrals and charged species.

The straightforward approach to this system is to directly solve the macroscopic equations, which can be achieved using various methods\cite{ballester_partially_2018, shang_computational_nodate}. For instance, within the finite volume method (FVM) framework, different numerical flux calculation methods such as the Roe solver\cite{roe_approximate_1981} or the HLL (Harten-Lax-van Leer)\cite{noauthor_upstream_nodate} solver can be used. Source terms can be incorporated into flux evaluation or split using Strang splitting methods\cite{ballester2018}. In astrophysics community, sometimes, the finite difference method is used to discretize these equations\cite{khomenko_three-dimensional_2018,popescu_braileanu_two-fluid_2019,felipe_magneto-acoustic_2010,martinez-gomez_simulations_2021,gonzalez-morales_mhdsts_2018}. While this approach is straightforward, it may not be convenient for constructing multiscale methods. In contrast, the other approach involves solving the underlying kinetic equations. This approach evaluates numerical flux by taking moments of distribution function at the cell interface. Although this approach is not straightforward, it is feasible to extend it to the whole flow regime by constructing multiscale methods. One representative work is the extension of the gas-kinetic scheme (GKS) for the continuum flow simulation to the unified GKS (UGKS) and discrete UGKS (DUGKS) for the multiscale flow simulations \cite{xu2010unified,guo2021progress,liu2021unified}.

Over the past decades, there has been systematic development of GKS for modeling compressible flows and magnetohydrodynamics \cite{xu_numerical_1993, xu2001,xu1999gas}. In the GKS formulation, a time-dependent gas distribution function is constructed at the cell interface from which hydrodynamic fluxes are obtained by taking moments of the distribution function. In smooth regions of the flow, the scheme accurately recovers either the Navier-Stokes or Euler solution depending on the mean relaxation time. Across discontinuities, artificial dissipation naturally arises from the free transport of particles on kinetic scales, enabling the scheme to capture shock waves. Basically, GKS inherently combines the merits of the Lax-Wendroff scheme and the upwind method, and makes a smooth transition  between these two limits according to the
local flow condition. The current research is to develop GKS for PIP system.

In this work, kinetic equations for electrons, ions, and neutrals based on the BGK-Maxwell\cite{liu2017} and AAP models\cite{andries2002} are developed. Through asymptotic analysis, the model reduces to the Euler equations for neutrals and ideal MHD equations for charged species in the continuum and strong magnetization limits. The model is solved numerically using an operator-splitting method for the particle transport and electromagnetic field evolution, while the fluid flow is solved by the gas-kinetic scheme (GKS) and the evolution of the electromagnetic field is computed with a wave-propagating finite volume scheme \cite{leveque_wave_1997}.
The coupled system is applied to many test cases, which include the Riemann problem to study the interaction of acoustic waves in the neutral species and MHD waves in the plasma. As a more complicated case, the Orszag-Tang vortex is tested at different ionization states to examine neutrals' effects on MHD wave interaction.

This paper is organized as follows:
In Section 2, the kinetic model for neutrals, electrons, and ions is introduced. After that, the asymptotic behaviors of the kinetic system are analyzed and the system reduces to the Euler equations and ideal MHD equations under the limiting condition.
In Section 3, the numerical methods are presented, including the gas kinetic scheme for PIP and the wave-propagation-based finite volume scheme for the Maxwell equations.
In Section 4, numerical examples are presented. Finally, in Section 5, conclusions are provided.

\section{Kinetic model and asymptotic behavior}
\label{model}
\subsection{BGK-Maxwell kinetic model}
For partially ionized plasma, the kinetic equation can be written as\cite{liu2017}:

\begin{equation}
    \begin{aligned}
    & \frac{\partial f_\alpha}{\partial t}+\boldsymbol{u}_\alpha \cdot \nabla_x f_\alpha+\boldsymbol{a}_\alpha \cdot \nabla_u f_\alpha= Q_{\alpha}, \\
    & \frac{\partial \boldsymbol{B}}{\partial t}+\nabla_x \times \boldsymbol{E}=0, \\
    & \frac{\partial \boldsymbol{E}}{\partial t}-\boldsymbol{c}^2 \nabla_x \times \boldsymbol{B}=-\frac{1}{\epsilon_0} \boldsymbol{J}, \\
    &\nabla_{\boldsymbol{x}} \cdot \boldsymbol{E}=\frac{q}{\epsilon_0},\\
    &\nabla_{\boldsymbol{x}} \cdot \boldsymbol{B}=0,
    \end{aligned}
\end{equation}
where $f_\alpha = f_\alpha(t, \boldsymbol{x}, \boldsymbol{u})$ is the distribution function for species $\alpha$ ( $\alpha=i$ for ion and $\alpha=e$ for electron, $\alpha=n$ for neutral) at space and time $(\boldsymbol{x},t)$ and microscopic translational velocity $\boldsymbol{u}$. $\boldsymbol{a}_\alpha$ is the Lorenz acceleration taking the form
$$
\boldsymbol{a}_\alpha=\frac{q_\alpha(\boldsymbol{E}+\boldsymbol{u}_\alpha \times \boldsymbol{B})}{m_\alpha}.
$$

For neutral species, $q_n=0$, thus $\boldsymbol{a}_n=0$. $Q_{\alpha}=\sum_{k=1}^{m}Q_{\alpha k}(f_\alpha, f_k)$ is the collision operator of species $\alpha$ between species $k$, where $m$ is total number of species in the system. In this work, $m=3$. In the Maxwell equations,  $\boldsymbol{E}$ and $\boldsymbol{B}$ are the electric field strength and magnetic induction, $\boldsymbol{c}$ is the speed of light, and $\epsilon_0$ is the vacuum permittivity. $n_\alpha$ is number density of species $\alpha$.In this work, the charged species in the system are just protons and electrons, then the electric current is $\boldsymbol{J}=e\left(n_i \boldsymbol{U}_i-n_e \boldsymbol{U}_e\right)$ and the charge density is $q=e(n_i-n_e)$, where $\boldsymbol{U}$ is macroscopic velocity and $e$ is the charge of a proton.

Collision between multiple species is modeled by the relaxation model by Andries, Aoki, and Perthanme\cite{andries2002}, which is,
\begin{equation}
    Q_\alpha=\frac{g_\alpha^M-f_\alpha}{\tau_\alpha},
\label{aap_operator}
\end{equation}
where $g_\alpha^{M}$ is a Maxwellian distribution,
\begin{equation}
    g_\alpha^M=\rho_\alpha\left(\frac{m_\alpha}{2 \pi k T_\alpha^*}\right)^{3 / 2} \exp \left(-\frac{m_\alpha}{2 k_B T_\alpha^*}\left(\boldsymbol{u}_\alpha-\boldsymbol{U}_\alpha^*\right)^2\right).
\end{equation}
and post-collision temperature and velocity are chosen as:
\begin{equation}
    \begin{aligned}
    \boldsymbol{U}_\alpha^* & =\boldsymbol{U}_\alpha+\frac{\tau_\alpha}{m_\alpha} \sum_{k=1}^N 2 \mu_{\alpha k} \chi_{\alpha k} n_k\left(\boldsymbol{U}_k-\boldsymbol{U}_\alpha\right), \\
    T_\alpha^* & =T_\alpha-\frac{m_\alpha}{3 k_B}\left(\boldsymbol{U}_\alpha^*-\boldsymbol{U}_\alpha\right)^2+\tau_\alpha \sum_{k=1}^N \frac{4 \mu_{\alpha k} \chi_{\alpha k} n_k}{m_\alpha+m_k}\left(T_k-T_\alpha+\frac{m_k}{3 k_B}\left(\boldsymbol{U}_k-\boldsymbol{U}_\alpha\right)^2\right),
    \end{aligned}
    \label{eq:aap U and E}
\end{equation}
where $\mu_{\alpha k} = m_{\alpha}m_k/(m_{\alpha}+m_k)$ is reduced mass, mean relaxation time $\tau_\alpha$ is determined by $1 / \tau_\alpha=\sum_{k=1}^m \chi_{\alpha k} n_{k} $, and interaction coefficient $\chi_{\alpha k}$ for hard sphere model is \cite{morse1963}:
\begin{equation*}
    \chi_{\alpha k}= \frac{4 \sqrt{\pi}}{3}\left(\frac{2 k_B T_{\alpha}}{m_\alpha}+\frac{2 k_B T_{k}}{m_k}\right)^{1 / 2}\left(\frac{d_\alpha+d_k}{2}\right)^2 .
\end{equation*}
In this above formula, $ d_\alpha, d_k $are the diameters of the particles and can be approximated by
$$(d_\alpha+d_k)^2 = \frac{1}{\sqrt{2}\pi (n_\alpha+n_k) \text{Kn} \text{L}},$$
where $\text{Kn}$ is Knudsen number, $\text{L}$ is reference length

To satisfy the divergence constraint, the Perfect Hyperbolic Maxwell equations (PHM) are used to reformulate the Maxwell equations as
\begin{align}
    & \frac{\partial \boldsymbol{E}}{\partial t}-c^2 \nabla_{\boldsymbol{x}} \times \boldsymbol{B}+\chi c^2 \nabla_{\boldsymbol{x}} \phi=-\frac{1}{\epsilon_0} \boldsymbol{J},\label{eq:PHM Amphere law}\\
    & \frac{\partial \boldsymbol{B}}{\partial t}+\nabla_{\boldsymbol{x}} \times \boldsymbol{E}+\gamma \nabla_{\boldsymbol{x}} \psi=0, \label{eq:PHM Faraday law}\\
    & \frac{1}{\chi} \frac{\partial \phi}{\partial t}+\nabla_{\boldsymbol{x}} \cdot \boldsymbol{E}=\frac{q}{{\epsilon_0}}, \label{eq:PHM E divergence}\\
    & \frac{\epsilon_0 \mu_0}{\gamma} \frac{\partial \psi}{\partial t}+\nabla_{\boldsymbol{x}} \cdot \boldsymbol{B}=0,\label{eq:PHM B divergence}
\end{align}
where $\phi,\psi$ are artificial correction potentials to accommodate divergence errors traveling at speed $\gamma c$ and $\chi c$\cite{munz_divergence_2000,munz_three-dimensional_2000}.

\subsection{Asymptotic analysis and fluid limit}
First, the kinetic governing equations are non-dimensionalized, resulting in a dimensionless kinetic system describing the behavior of electrons, protons, and neutrals.
The Chapman-Enskog method can be applied to derive a three-fluid model from the kinetic system. Within this three-fluid model, the electron and ion fluids constitute a two-fluid subsystem.
By analyzing an additional small parameter, the ideal Magnetohydrodynamic (MHD) limit is obtained from the two-fluid subsystem. In this limit, the electron and ion fluids are combined into a single conducting fluid. The ideal MHD equations describe the dynamics of this conducting fluid.
The end result is a coupled system of equations for the ideal MHD conducting fluid and the separate neutral fluid. This coupled system describes the ideal partially ionized plasma in the MHD limit in Eq.\eqref{eq:Euler-MHD}.

The reference quantities are defined as follows:
$$
u_0 = \sqrt{2k_BT_0/m_i}, \rho_0 = m_0 n_0, E_0 = B_0 u_0, a_0 = eB_0u_0/m_0, f_0=m_0n_0/u_0^3, c_0 = u_0\sqrt{\gamma/2},
$$
where charateristic velocity $u_0$ is thermal velocity of ions, and $m_0=m_i$ is mass of ions. $\rho_0$ is reference density, $E_0$ and $B_0$ are reference electric and magnetic field strength, $a_0$ is reference acceleration, $f_0$ is reference distribution function, $c_0$ is reference sound speed and $\gamma$ is specific heat. Then variables are non-dimensionalized as:
\begin{equation*}
    \begin{aligned}
    &\hat{x}=\frac{x}{l_0}, \hat{\boldsymbol{u}}=\frac{\boldsymbol{u}}{u_0}, \hat{t}=\frac{u_0}{l_0} t, \hat{m}=\frac{m}{m_0}, \hat{n}=\frac{n}{n_0},\hat{\mathscr{E}}=\frac{\mathscr{E}}{m_i n_0 u_0^2}, \hat{f}=\frac{u_0^3}{m_0 n_0} f, \hat{\boldsymbol{B}}=\frac{\boldsymbol{B}}{B_0}, \\
    &\hat{\boldsymbol{E}}=\frac{\boldsymbol{E}}{B_0 u_0}, \hat{\boldsymbol{J}}=\frac{\boldsymbol{J}}{en_0 u_0},\hat{\lambda}_D=\frac{\lambda_D}{r_{Li}} = \sqrt{\frac{\epsilon_0m_0u_0^2}{ne^2}}\frac{eB_0}{m_0u_0}, \hat{r_{L}} = \frac{r_{L}}{l_0} = \frac{m_0u_0}{eB_0l_0}.
    \end{aligned}
\end{equation*}

Inserting the normalized variables into the BGK-Maxwell system, the following dimensionless BGK-Maxwell system is obtained:

\begin{equation}
    \begin{aligned}
    &\frac{\partial \tilde{f}_{\alpha}}{\partial \tilde{t}}+\tilde{\boldsymbol{u}_\alpha} \cdot \nabla_{\tilde{x}} \tilde{f}_{\alpha}+\frac{q_{\alpha}(\tilde{\boldsymbol{E}}+\tilde{\boldsymbol{u}_\alpha }\times\tilde{\boldsymbol{B}})}{m_\alpha\tilde{r}}  \cdot \nabla_{\tilde{u}} \tilde{f}_{\alpha}=\frac{\tilde{g}_{\alpha}^M-\tilde{f}_{\alpha}}{\tilde{\tau}_{\alpha}}, \\
    &\frac{\partial \tilde{\boldsymbol{B}}}{\partial \tilde{t}}+\nabla_{\tilde{x}} \times \tilde{\boldsymbol{E}}=0, \\
    &\frac{\partial \tilde{\boldsymbol{E}}}{\partial \tilde{t}}-\tilde{c}^{2} \nabla_{\tilde{x}} \times \tilde{\boldsymbol{B}}=-\frac{1}{\tilde{\lambda}_{D}^{2} \tilde{r_L}} \boldsymbol{J},\\
    &\nabla_{\tilde{\boldsymbol{x}}} \cdot \tilde{\boldsymbol{E}}=\frac{\tilde{n_i}-\tilde{n_e}}{\tilde{\lambda_D}^2 \tilde{r_{L}}}, \quad \nabla_{\tilde{\boldsymbol{x}}} \cdot \tilde{\boldsymbol{B}}=0,
    \end{aligned}
    \label{BGKnondim}
\end{equation}
where $\tilde{\lambda_D}$ is normalized Debye length, $\tilde{r}_{L}$ is normalized larmor radius. For simplicity, in the following of this paper, all the hats for nondimensionalized variables are omitted.

Zeroth-order asymptotic solution of $f$ based on Chapman-Enskog expansion is
$
    f=g+O\left(\tau^{1}\right)
$\cite{xu2001}. Substitute the solution into BGK equation and take moments, a three-fluid system composed of neutral species, ions, and electrons is obtained,
\begin{equation}
\begin{aligned}
& \partial_t \rho_n+\nabla_{\boldsymbol{x}} \cdot\left(\rho_n \boldsymbol{U}_n\right)=0, \\
& \partial_t\left(\rho_n \boldsymbol{U}_n\right)+\nabla_{\boldsymbol{x}} \cdot\left(\rho_n \boldsymbol{U}_n \boldsymbol{U}_n+p_n \mathbb{I}\right)=S_n, \\
& \partial_t \mathscr{E}_n+\nabla_{\boldsymbol{x}} \cdot\left(\left(\mathscr{E}_n+p_n\right) \boldsymbol{U}_n\right)=Q_n,\\
& \partial_t \rho_i+\nabla_{\boldsymbol{x}} \cdot\left(\rho_i \boldsymbol{U}_i\right)=0, \\
& \partial_t\left(\rho_i \boldsymbol{U}_i\right)+\nabla_{\boldsymbol{x}} \cdot\left(\rho_i \boldsymbol{U}_i \boldsymbol{U}_i+p_i \mathbb{I}\right)=\frac{q_in_i}{r_{L}}\left(\boldsymbol{E}+\boldsymbol{U}_i \times \boldsymbol{B}\right)+S_i, \\
& \partial_t \mathscr{E}_i+\nabla_{\boldsymbol{x}} \cdot\left(\left(\mathscr{E}_i+p_i\right) \boldsymbol{U}_i\right)=\frac{q_in_i}{r_{L}} \boldsymbol{U}_i \cdot \boldsymbol{E}+Q_i,\\
& \partial_t \rho_e+\nabla_{\boldsymbol{x}} \cdot\left(\rho_e \boldsymbol{U}_e\right)=0, \\
& \partial_t\left(\rho_e \boldsymbol{U}_e\right)+\nabla_{\boldsymbol{x}} \cdot\left(\rho_e \boldsymbol{U}_e \boldsymbol{U}_e+p_e \mathbb{I}\right)=\frac{q_en_e}{r_{L}}\left(\boldsymbol{E}+\boldsymbol{U}_e \times \boldsymbol{B}\right)+S_e, \\
& \partial_t \mathscr{E}_e+\nabla_{\boldsymbol{x}} \cdot\left(\left(\mathscr{E}_e+p_e\right) \boldsymbol{U}_e\right)=\frac{q_en_e}{r_{L}} \boldsymbol{U}_e \cdot\boldsymbol{E}+Q_e, \\
&\frac{\partial {\boldsymbol{B}}}{\partial {t}}+\nabla_{{x}} \times {\boldsymbol{E}}=0, \quad \frac{\partial {\boldsymbol{E}}}{\partial {t}}-{c}^{2} \nabla_{{x}} \times {\boldsymbol{B}}=-\frac{1}{{\lambda}_{D}^{2} {r_L}} \boldsymbol{J},\\
&\nabla_{{\boldsymbol{x}}} \cdot {\boldsymbol{E}}=\frac{{n_i}-{n_e}}{{\lambda_D}^2 {r_{L}}}, \quad \nabla_{{\boldsymbol{x}}} \cdot {\boldsymbol{B}}=0,
\label{threefluid}
\end{aligned}
\end{equation}
where $S_\alpha$ and $Q_\alpha$ are momentum and energy exchange between species $\alpha$ and other species in the system.

Except the first three equations for the neutral gas, the rest system in Eq.\eqref{threefluid} forms an ion-electron two-fluid subsystem, based on which Hall-effect MHD and ideal MHD can be derived\cite{liu2017,shen_magnetohydrodynamic_2018}.

Define center-of-mass velocity as
$$
\boldsymbol{U}=\frac{m_i \boldsymbol{U}_i+m_e \boldsymbol{U}_e}{m_i+m_e},
$$
Denote mass ratio $\epsilon = m_e/m_i$,
$(1+\epsilon)\boldsymbol{U} = \boldsymbol{U}_i + \epsilon \boldsymbol{U}_e $.
For $\mathcal{O}(\epsilon^0)$ balance,
$$
\boldsymbol{U}_i = \boldsymbol{U} \quad \text{and}\quad \boldsymbol{U}_e = \boldsymbol{U} - \boldsymbol{U}_i + \boldsymbol{U}_e  = \boldsymbol{U} - \frac{\boldsymbol{J}}{ne},
$$

Substituting the above approximation into an electron momentum equation in the two-fluid subsystem,
$$
\boldsymbol{E}+\boldsymbol{U} \times \boldsymbol{B}=\frac{2 m_i m_e v_{i e}}{\left(m_i+m_e\right) n_e e^2} \boldsymbol{J}+\frac{r_{L} }{n_e e} \boldsymbol{J} \times \boldsymbol{B}+\frac{r_{L}}{n_e e} \partial_t\left(\rho_e \boldsymbol{U}_e\right)+\frac{r_{L}}{n_e e} \nabla_{\boldsymbol{x}} \cdot\left(\rho_e \boldsymbol{U}_e \boldsymbol{U}_e+p_e \mathbb{I}\right).
$$

The right-hand side of the above equation contains four terms: the first term represents electric resistivity, the second term corresponds to the Hall effect, and the last two terms describe the effects of electron inertia and pressure. In the limit $\epsilon \rightarrow 0$, the electron's inertia can be negected and the electron momentum equation gives the generalized Ohm's law
\begin{equation}
\boldsymbol{E}+\boldsymbol{U} \times \boldsymbol{B}=\frac{1}{\sigma} \boldsymbol{J}+\frac{r_{L}}{n_e e} \boldsymbol{J} \times \boldsymbol{B}+\frac{r_{L}}{n_e e} \nabla_{\mathrm{x}} p_e.
\label{eq:general ohm}
\end{equation}

For non-relativistic flows, the displacement current is negligible in view of,
$$
\frac{1}{c^2}|\frac{\partial \boldsymbol{E}}{\partial t}| \sim \frac{U^2}{c^2} \frac{B}{L} \ll |\nabla\times \boldsymbol{B}| \sim \frac{B}{L},
$$
where $U$ and $L$ is the characteristic macroscale velocity and spatial length of plasma, and $U^2/c^2\ll 1$.
When $\lambda_D \sim c^{-1} \rightarrow 0$ (i.e., $\lambda c^{-1}=1$), the Ampère's law then becomes
$$
\boldsymbol{J}=r_{L}  \nabla_{\boldsymbol{x}} \times \boldsymbol{B}+\mathcal{O}\left(U^2 / c^2\right).
$$
The above low-frequency Ampère's law indicates that $\nabla \cdot \boldsymbol{J}=0$, and therefore in this regime, the plasma is quasi-neutral, namely $n_i \approx n_e$.

In such a regime, the two fluid equations reduce to one fluid Hall-MHD equation. The Hall term and the electron pressure term are on the order of Larmor radius. Then Hall-MHD can be written as,
$$
\begin{aligned}
& \partial_t \rho+\nabla_{\boldsymbol{x}} \cdot(\rho \boldsymbol{U})=0, \\
& \partial_t(\rho \boldsymbol{U})+\nabla_{\boldsymbol{x}} \cdot\left(\rho \boldsymbol{U U}+p_i \mathbb{I}\right)=\frac{\rho_i}{m_i r_{L}}(\boldsymbol{E}+\boldsymbol{U} \times \boldsymbol{B}), \\
& \boldsymbol{E}+\boldsymbol{U} \times \boldsymbol{B}=\frac{1}{\sigma} \boldsymbol{J}+\frac{r_{L}}{n_e e} \boldsymbol{J} \times \boldsymbol{B}+\frac{r_{L}}{n_e e} \nabla_{\boldsymbol{x}} p_e, \\
& \partial_t \mathscr{E}_\alpha+\nabla_{\boldsymbol{x}} \cdot\left(\left(\mathscr{E}_\alpha+p_\alpha\right) \boldsymbol{U}_\alpha\right)=\frac{1}{r_{L}} \boldsymbol{J}\cdot \boldsymbol{E}, \\
& \partial_t \boldsymbol{B}+\nabla_{\boldsymbol{x}} \times \boldsymbol{E}=0, \\
& \boldsymbol{J}=r_{L} \nabla_{\boldsymbol{x}} \times \boldsymbol{B}.
\end{aligned}
$$

In the limit $r_{L} \rightarrow 0$, the Hall current and electron pressure in Eq.\eqref{eq:general ohm} are negligible. And ignoring collisions between electrons and protons i.e. $\nu_{ie}=0$, electric resistivity thus can be dropped. Then the generalized Ohm's law reduces to the ideal Ohm's law,
$$
\boldsymbol{E}+\boldsymbol{U} \times \boldsymbol{B}=0.
$$
So combining electron and ion momentum equations, ideal MHD equation can be written as,
$$
\begin{aligned}
& \partial_t \rho+\nabla_{\boldsymbol{x}} \cdot(\rho \boldsymbol{U})=0, \\
& \partial_t(\rho \boldsymbol{U})+\nabla_{\boldsymbol{x}} \cdot(\rho \boldsymbol{U} \boldsymbol{U}+p \mathbb{I})= (\nabla\times\boldsymbol{B}) \times \boldsymbol{B}, \\
& \partial_t \mathscr{E}+\nabla_{\boldsymbol{x}} \cdot((\mathscr{E}+p) \boldsymbol{U})= (\nabla\times\boldsymbol{B}) \cdot\boldsymbol{E},\\
& \partial_t \boldsymbol{B}+\nabla_{\boldsymbol{x}} \times(\boldsymbol{U} \times \boldsymbol{B})=0.\\
\end{aligned}
$$
In this limit, the three-fluid system in Eq.\eqref{threefluid} turns to a neutral and ideal MHD two-fluid system.

\begin{figure}[H]
\centering
\includegraphics[width=0.99\textwidth]{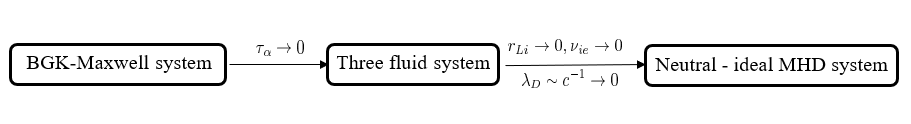}
\caption{Asymptotic behavior of BGK-Maxwell system}
\label{fig:asymptotic behavior of BGK-Maxwell}
\end{figure}

In summary, as shown in Fig.\ref{fig:asymptotic behavior of BGK-Maxwell} in the limit of $r_L\rightarrow 0, \lambda_D \sim c^{-1} \rightarrow 0, m_e/m_i \rightarrow 0, \tau \rightarrow 0,\nu_{ie}\rightarrow 0$, the kinetic system in Eq.\eqref{BGKnondim} can describe the ideal partially ionized plasma in Eq.\eqref{eq:Euler-MHD}.

\section{Numerical method}
\label{numerical}

\subsection{General framework}
In the framework of the FVM, the cell averaged conservative variables for species $\alpha$ is $(\boldsymbol{W}_{\alpha})_i = ((\rho_{\alpha})_i, (\rho_{\alpha}\boldsymbol{U}_{\alpha})_i, (\rho_{\alpha}\mathscr{E}_{\alpha})_i)$ on a physical cell $\Omega_i$ are defined as
$$
(\boldsymbol{W}_{\alpha})_i = \frac{1}{|\Omega_i|}\int_{\Omega_i}\boldsymbol{W}_{\alpha}(\boldsymbol{x})\mathrm{d}\boldsymbol{x},
$$
where $|\Omega_i|$ is the volume of cell $\Omega_i$. For a discretized time step $\Delta t=t^{n+1}-t^n$, the evolution of $(\boldsymbol{W}_{\alpha})_i$ is
\begin{equation}
(\boldsymbol{W}_{\alpha})_i^{n+1} = (\boldsymbol{W}_{\alpha})_i^n - \frac{\Delta t}{|\Omega_i|}\sum_{s\in\partial\Omega_i}|l_s|(\mathscr{F}_{\boldsymbol{W}_\alpha})_s + \frac{\Delta t}{\tau_\alpha}(\bar{(\boldsymbol{W}}_{\alpha})_i^n - (\boldsymbol{W}_{\alpha})_i^n)+\Delta t(\boldsymbol{S}_{\alpha})_i^{n+1},
\label{eq:FVM discretization}
\end{equation}
where $l_s\in \partial \Omega_i$ is the cell interface with center $\boldsymbol{x}_s$ and outer unit normal vector $\boldsymbol{n}_{s}$. $|l_s|$ is the area of the cell interface. $(\bar{\boldsymbol{W}}_{\alpha})_i = ((\rho_{\alpha})_i, (\rho_{\alpha}\bar{\boldsymbol{U}}_{\alpha})_i, (\rho_{\alpha}\bar{\mathscr{E}}_{\alpha})_i)$
where $(\bar{\boldsymbol{U}}_{\alpha})_i$ and $(\bar{\mathscr{E}}_{\alpha})_i$ are post-collision velocity and energy in AAP model as Eq.\eqref{eq:aap U and E}. $(\boldsymbol{S}_{\alpha})_i$ is source term due to electromagnetic force. The numerical flux across interface $(\mathscr{F}_{\boldsymbol{W}_\alpha})_s$ can be evaluated from distribution function at the interface,
\begin{equation}
(\mathscr{F}_{\boldsymbol{W}_\alpha})_s = \frac{1}{\Delta t}\int_{t^n}^{t^{n+1}}\boldsymbol{u}\cdot \boldsymbol{n}_sf_\alpha(\boldsymbol{x}_{s},\boldsymbol{u},\boldsymbol{\xi},t)\boldsymbol{\Psi} \mathrm{d}\boldsymbol{\Xi}\mathrm{d}t,
\label{eq:FVM macroscopic flux}
\end{equation}
where $\boldsymbol{\Psi}=(1,\boldsymbol{u},\frac{1}{2}(\boldsymbol{u}^2+\boldsymbol{\xi}^2))$ is the conservative moments of distribution functions with $\boldsymbol{\xi}=(\xi_1,\xi_2,\cdots,\xi_n)$ the internal degree of freedom. $\mathrm{d}\boldsymbol{\Xi}=\mathrm{d}\boldsymbol{u}d\boldsymbol{\xi}$ is the volume element in the phase space. The evaluation of the distribution function will be presented in section \ref{Gas-kineitc scheme}.

To be specific, the elementwise equations in Eq.\eqref{eq:FVM discretization} are

\begin{equation}
\begin{aligned}
\left(\rho_\alpha\right)_i^{n+1}= & \left(\rho_\alpha\right)_i^n-\frac{\Delta t}{\left|\Omega_{i}\right|} \sum_{s \in \partial \Omega_{{i}}}\left|l_s\right| (\mathscr{F}_{\rho_\alpha})_s, \\
\left(\boldsymbol{\rho}_\alpha \boldsymbol{U}_\alpha\right)_i^{n+1}= & \left(\boldsymbol{\rho}_\alpha \boldsymbol{U}_\alpha\right)_i^n-\frac{\Delta t}{\left|\Omega_{i}\right|} \sum_{s \in \partial \Omega_{{i}}}\left|l_s\right| (\mathscr{F}_{(\rho\boldsymbol{U})_\alpha})_s\\
& +\frac{\Delta t}{\tau_\alpha}\left(\rho_\alpha^n \bar{\boldsymbol{U}^n}{ }_\alpha-\rho_\alpha^n \boldsymbol{U}_\alpha^n\right)+\frac{\Delta t}{r_{L_i}} n_\alpha^{n+1}\left(\boldsymbol{E}^{n+1}+\boldsymbol{U}_\alpha^{n+1} \times \boldsymbol{B}^{n+1}\right), \\
\left(\boldsymbol{\rho}_\alpha \mathscr{E}_\alpha\right)_i^{n+1}= & \left(\boldsymbol{\rho}_\alpha \mathscr{E}_\alpha\right)_i^n-\frac{\Delta t}{\left|\Omega_{i}\right|} \sum_{s \in \partial \Omega_{i}}\left|l_s\right| (\mathscr{F}_{(\rho\mathscr{E})_\alpha})_s \\
& +\frac{\Delta t}{\tau_\alpha}\left(\rho_\alpha^n \bar{\mathscr{E}^n}{ }_\alpha-\rho_\alpha^n \mathscr{E}_\alpha^n\right)+\frac{\Delta t}{r_{L_i}} n_\alpha^{n+1} \boldsymbol{U}_\alpha^{n+1} \cdot \boldsymbol{E}^{n+1} .
\end{aligned}
\label{eq:fvm fluid elementwise}
\end{equation}

In Eq.\eqref{eq:fvm fluid elementwise}, the source term due to cross-species momentum and energy exchange will be evaluated by the operator splitting method. In the Euler limit,$\tau_\alpha\rightarrow 0$, $(\bar{\boldsymbol{W}_\alpha})\rightarrow\boldsymbol{W}_\alpha)$ is obtained at every time step. Lorentz source term is split and coupled with source terms in PHM so as to get coupled evolution between fluid species and electromagnetic field. The interaction equations can be solve by Crank-Nicolson scheme introduced in Section \ref{interaction equation}.

The cell averaged quantities $\boldsymbol{Q}_i$ for electromagnetic variables $\boldsymbol{Q} = \left(E_{x}, E_{y}, E_{z}, B_{x}, B_{y}, B_{z}, \phi, \psi\right)$ in a cell are defined as
$$\boldsymbol{Q}_i =  \frac{1}{|\Omega_i|}\int_{\Omega_i}\boldsymbol{Q}(\boldsymbol{x})\mathrm{d}\boldsymbol{x}, $$
where $(\mathscr{F}_{\boldsymbol{Q}})_s$ is numerical flux across the cell interface $l_s$ which will be presented in section \ref{sec:phm}.
The time evolution formula is
$$
\boldsymbol{Q}_i^{n+1}=\boldsymbol{Q}_i^n+\frac{\Delta t}{\left|\Omega_{i}\right|} \sum_{s \in \partial \Omega_{i}}\left|l_i\right| (\mathscr{F}_{\boldsymbol{Q}})_s+\Delta t (\boldsymbol{S}_{\boldsymbol{Q}})_i^{n+1},
$$
where $(\mathscr{F}_{\boldsymbol{Q}})_s$ is numerical flux across a cell interface, which will be presented in Section \ref{sec:phm}. $\boldsymbol{S}_{\boldsymbol{Q}}$ are sources terms in PHM equations. The componentwise equations are:
$$
\begin{aligned}
\boldsymbol{E}_i^{n+1} & =\boldsymbol{E}_i^n+\frac{\Delta t}{\left|\Omega_{ i}\right|} \sum_{s \in \partial \Omega_{i}}\left|l_s\right| (\mathscr{F}_{\boldsymbol{E}})_s-\frac{\Delta t}{\lambda_D^2 r_{L}}\left(n_i^{n+1} \boldsymbol{U}_i^{n+1}-n_e^{n+1} \boldsymbol{U}_e^{n+1}\right), \\
\boldsymbol{B}_i^{n+1} & =\boldsymbol{B}_i^n+\frac{\Delta t}{\left|\Omega_{ i}\right|} \sum_{s \in \partial \Omega_{i}}\left|l_s\right| (\mathscr{F}_{\boldsymbol{B}})_s, \\
\phi_i^{n+1} & =\phi_i^n+\frac{\Delta t}{\left|\Omega_{ i}\right|} \sum_{s \in \partial \Omega_{{i}}}\left|l_s\right| (\mathscr{F}_{\phi})_s+\frac{\Delta t \chi}{\lambda_D^2 r_{L}}\left(n_i^{n+1}-n_e^{n+1}\right), \\
\psi_i^{n+1} & =\psi_i^n+\frac{\Delta t}{\left|\Omega_{i}\right|} \sum_{s \in \partial \Omega_{i}}\left|l_s\right| (\mathscr{F}_{\psi})_s.
\end{aligned}
$$

General numerical steps are listed as follows:
\begin{enumerate}
    \item Update conservative variable $\boldsymbol{W}_\alpha^{n}$ to $\boldsymbol{W}_\alpha^{*}$ considering net flux without force term across the cell interface.
    \item Update conservative variable $\boldsymbol{W}_\alpha^{*}$ to $\boldsymbol{W}_\alpha^{**}$ considering momentum and energy exchange between different species.
    \item Update electromagnetic field $\boldsymbol{E}^{n} \rightarrow \boldsymbol{E}^{*}$, $\boldsymbol{B}^{n} \rightarrow \boldsymbol{B}^{n+1}$, ${\phi}^{n} \rightarrow {\phi}^{*}$ and ${\psi}^{n} \rightarrow {\psi}^{n+1}$ by net flux across physical cell interface.
    \item Incorporate the interaction between electromagnetic field and charged species $\boldsymbol{W}_\alpha^{**}$ to $\boldsymbol{W}_\alpha^{n+1}$,$\boldsymbol{E}^{*} \rightarrow \boldsymbol{E}^{n+1}$,${\phi}^{*} \rightarrow {\phi}^{n+1}$.
\end{enumerate}
After four steps, all variables are evolved from $t^n$ to $t^{n+1}$.

\subsection{Numerical flux of conservative variables}
\label{Gas-kineitc scheme}
In this section, the solution distribution function at a cell interface is constructed so as to calculate numerical flux in Eq.\eqref{eq:FVM macroscopic flux} in the two-dimensional (2D) case. The 2D BGK equation without force term can be written as
$$
     \frac{\partial f}{\partial t}+\boldsymbol{u} \cdot \nabla_x f = \frac{g-f}{\tau},
$$
where $\boldsymbol{u} = (u,v)$. For simplicity, species subscript $\alpha$ is omitted here.
To get the distribution function at a cell interface in Eq.\eqref{eq:FVM macroscopic flux}, the time evolution solution at the interface can be written as,
\begin{equation}
\begin{aligned}
    f\left({x}_{i+1/2,j}, {y}_{i+1/2,j},t, {u}, v, \boldsymbol{\xi}\right)=&\frac{1}{\tau} \int_{0}^{t} g\left({x}^{\prime}, y^{\prime}, t^{\prime}, {u}, v, \boldsymbol{\xi}\right) e^{-\left(t-t^{\prime}\right) / \tau} \mathrm{d} t^{\prime}\\
    &+e^{-t / \tau_n} f_{0}\left({x}_{i+1/2,j}-{u} t , y_{i+1/2,j}-vt\right),
    \label{eq:BGKsoln}
\end{aligned}
\end{equation}
where $(x_{i+1/2,j} = 0, y_{i+1/2,j}) = (0,0)$ is point on the interface. $\boldsymbol{\xi}=(w,\xi_1,\xi_2,\cdots,\xi_n)$ is the internal degree of freedom. $f_0$ is the initial gas distribution function at $t=0$, and $g$ is equilibrium distribution at $({x}^{'},y^{'},t^{'})$. $(x^{'}, y^{'} ) = (x_{i+1/2,j}-u(t-t^{'}), y_{i+1/2,j} - v(t-t^{'}))$ are the particle trajectory.  $\tau$ is mean relaxation time between successive collisions. $\tau_n$ is the numerical collision time. For the inviscid flow, the collision time $\tau_n$ is
$$
\tau_n=C_1 \Delta t+C_2\left|\frac{p_l-p_r}{p_l+p_r}\right| \Delta t,
$$
where $C_1=0.01$ and $C_2=5$. $p_l$ and $p_r$ denote the pressure on the left and right sides of the cell interface. The inclusion of the pressure jump term is to increase the non-equilibrium transport mechanism in the flux function to mimic the physical process in the shock layer\cite{xu2001}.

For the inviscid flow computation, the initial distribution at the cell interface can be reconstructed from cell average value as,
\begin{equation*}
    f_{0}= \begin{cases}g^{l}\left(1+a^{l} x + b^{l}y\right), & x \leq 0, \\ g^{r}\left(1+a^{r}x +b^{r}y\right), & x \geq 0,\end{cases}
\end{equation*}
$g^l, g^r$ are the equilibrium states at left and right cells. $a^l, a^r, b^l, b^r$ is the spatial slope of the initial distribution along the x and y directions at the left and right cells, i.e
$$
(a^l, a^r, b^l, b^r) = (\frac{\partial g^l}{\partial x}, \frac{\partial g^r}{\partial x}, \frac{\partial g^l}{\partial y}, \frac{\partial g^r}{\partial y}).
$$
The slope of the Maxwellian distribution can be evaluated as,
\begin{align*}
    \frac{\partial g}{\partial x} &= \frac{\partial g}{\partial \rho}\frac{\partial \rho}{\partial x} +
 \frac{\partial g}{\partial U}\frac{\partial U}{\partial x} + \frac{\partial g}{\partial V}\frac{\partial V}{\partial x}  +\frac{\partial g}{\partial \lambda}\frac{\partial \lambda}{\partial x}\\
 &= g(a_1 + a_2u + a_3v + \frac{1}{2}a_4(u^2+v^2+\boldsymbol{\xi}^2)),
\end{align*}
where the specific form of $a_1-a_4$ can be found in paper \cite{xu2001}.

Equilibrium distribution can be approximated as
\begin{equation*}
    g=g_{0}\left[1+(1-\mathrm{H}[x]) \bar{a}^{l} x+\mathrm{H}[x] \bar{a}^{r} x+\bar{b}y+\bar{A} t\right],
\end{equation*}
where $g_{0}$ is the equilibrium state at the interface,
\begin{equation*}
    \int g_{0} \psi d \Xi=w_{0}=\int_{u>0} \int g^{l} \psi \mathrm{d} \Xi+\int_{u<0} \int g^{r} \psi \mathrm{d} \Xi,
\end{equation*}
where $H(x)$ is Heaviside function. $\bar a^l, \bar a^r$ are microscopic spatial slopes of equilibrium distributions, $\bar A$ is the temporal slope, which can be obtained from macroscopic variables. $g^l$ and $g^r$ is the Maxwellian distribution at the left and right of the cell interface.

Finally, the gas distribution function $f$ at a cell interface can be written as,
\begin{equation}
    \begin{aligned}
    &f\left({x}_{i+1/2,j}, {y}_{i+1/2,j},t, {u}, v, \boldsymbol{\xi}\right)\\=&\left(1-e^{-t / \tau_n}\right) g_{0}
    +t e^{-t / \tau_n}\left(\bar{a}^{l} \mathrm{H}[u]+\bar{a}^{r}(1-\mathrm{H}[u]) + \bar{b}v \right) u g_{0}
    +t \bar{A} g_{0} \\
    &+e^{-t / \tau_n}\left((1-a^{l}ut) \mathrm{H}[u] g^{l}
    +\left(1-a^{r}ut \right)(1-\mathrm{H}[u]) g^{r}\right) .
    \end{aligned}
    \label{eq:BGK approximate soln}
\end{equation}
Substituting the above formula into Eq.\eqref{eq:FVM macroscopic flux}, the numerical flux can be calculated.

\subsection{Numerical flux of electromagnetic variables}
\label{sec:phm}
1D numerical flux is illustrated here, for 2D or 3D problems, simply rotating coordinate can be used to get flux in another direction. The general expression for 1D PHM system is:
\begin{equation}
    \frac{\partial \boldsymbol{q}}{\partial t}+\boldsymbol{A}_{1} \frac{\partial \boldsymbol{q}}{\partial x}=\boldsymbol{s},
    \label{eq:2D hyperbolic system}
\end{equation}
where
\begin{equation*}
    \boldsymbol{A}_{1}=\left(\begin{array}{cccccccc}
    0 & 0 & 0 & 0 & 0 & 0 & c^{2} \chi & 0 \\
    0 & 0 & 0 & 0 & 0 & c^{2} & 0 & 0 \\
    0 & 0 & 0 & 0 & -c^{2} & 0 & 0 & 0 \\
    0 & 0 & 0 & 0 & 0 & 0 & 0 & \gamma \\
    0 & 0 & -1 & 0 & 0 & 0 & 0 & 0 \\
    0 & 1 & 0 & 0 & 0 & 0 & 0 & 0 \\
    \chi & 0 & 0 & 0 & 0 & 0 & 0 & 0 \\
    0 & 0 & 0 & c^{2} \gamma & 0 & 0 & 0 & 0
    \end{array}\right).
\end{equation*}

The numerical flux across interface $(i-1/2,j)$ is \cite{leveque_wave_1997}:
\begin{equation}
    \begin{aligned}
    [\mathscr{F}_{\boldsymbol{Q}}]_{i-1 / 2, j}=& \frac{1}{2}\left(\boldsymbol{A}_{1} \boldsymbol{Q}_{i, j}+\boldsymbol{A}_{1} \boldsymbol{Q}_{i-1, j}\right)-\frac{1}{2}\left(\boldsymbol{A}_{1}^{+} \Delta \boldsymbol{Q}_{i-1 / 2}-\boldsymbol{A}_{1}^{-} \Delta \boldsymbol{Q}_{i-1 / 2}\right) \\
    &+\frac{1}{2} \sum_{p} \operatorname{sign}\left(\lambda_{i-1 / 2, j}^{p}\right)\left(1-\frac{\Delta t}{\Delta x} |\lambda_{i-1 / 2, j}^{p}|\right) \mathcal{L}_{1, i-1 / 2, j}^{p} \Phi\left(\theta_{1, i-/ 2, j}^{p}\right),
    \end{aligned}
    \label{eq:PHM flux}
\end{equation}
where $\mathcal{A}_{1}^{+} = R_1\Lambda^+R_1^{-1}$ and $\mathcal{A}_{1}^{-} = R_1\Lambda^-R_1^{-1}$. $R_1$ is the matrix composed of right eigenvectors of $A_1$, and $\Lambda^{+} = diag((\lambda^1)^{+},(\lambda^2)^{+},\cdots,(\lambda^8)^{+})$ with $\lambda^{+} = max(\lambda,0)$ and $\Lambda^{-} = diag((\lambda^1)^{-},(\lambda^2)^{-},\cdots,(\lambda^8)^{-})$ with $\lambda^{-} = min(\lambda,0)$. $\lambda^p$ is the $p$th eigenvalue of $A_1$. Besides,$\Delta \boldsymbol{Q}_{i-1 / 2} = \boldsymbol{Q}_{i+1}-\boldsymbol{Q}_{i}$. The flux slope in Eq.\eqref{eq:PHM flux} is
\begin{equation*}
    \mathcal{L}_{1, i-1 / 2, j}^{p}=\boldsymbol{l}_{1, i-1 / 2, j}^{p} \cdot\left(\boldsymbol{f}_{1, i, j}-\boldsymbol{f}_{1, i-1, j}\right) \boldsymbol{r}_{1, i-1 / 2, j}^{p},
\end{equation*}
where $\boldsymbol{l}$ and $\boldsymbol{r}$ are the left and right eigenvectors corresponding to eigenvalue $\lambda^p$. $\boldsymbol{f}$ is flux function in Eq.\eqref{eq:2D hyperbolic system}.
The limiter function $\Phi(\theta)$ is,
\begin{align*}
    \theta_{1, i-1 / 2, j}^{p} \equiv \frac{\mathcal{L}_{1, I-1 / 2, j}^{p} \cdot \mathcal{L}_{1, i-1 / 2, j}^{p}}{\mathcal{L}_{1, i-1 / 2, j}^{p} \cdot \mathcal{L}_{1, i-1 / 2, j}^{p}},
    \phi(\theta)=\max (0, \min ((1+\theta) / 2,2,2 \theta)),
\end{align*}
where $I=i-1$ if $\lambda_{i-1/2}^p>0$ and $I=i+1$ if $\lambda^p_{i-1/2}<0$. With the limiters, the scheme is second-order accurate in smooth region and first-order at or near  discontinuity.

\subsection{Electromagnetic field on fluid evolution }
\label{interaction equation}

The fluid evolution due to the interaction with electromagnetic field is
\begin{equation*}
    \begin{aligned}
    & \frac{\partial (\rho_i \boldsymbol{U}_i)}{\partial t}=\frac{e n_i}{r_L}\left(\boldsymbol{E}+\boldsymbol{U}_i \times \boldsymbol{B}\right), \\
    & \frac{\partial (\rho_e \boldsymbol{U}_e)}{\partial t}=-\frac{e n_e}{r_L}\left(\boldsymbol{E}+\boldsymbol{U}_e \times \boldsymbol{B}\right), \\
    & \frac{\partial \boldsymbol{E}}{\partial t}=-\frac{e}{\lambda_D^2 r_L}\left(\boldsymbol{U}_i-\boldsymbol{U}_e\right),\\
    &\frac{1}{\chi} \frac{\partial \phi}{\partial t}=\frac{{n_i}-{n_e}}{{\lambda_D}^2 {r_{L}}}.
    \end{aligned}
\end{equation*}
The above equations can be discretized by the Crank-Nicolson scheme,
\begin{equation*}
    \begin{aligned}
        &\boldsymbol{U}_i^{n+1} - \boldsymbol{U}_i^{**} = \frac{e\Delta t}{m_ir_L}(\frac{\boldsymbol{E}^{n+1} + \boldsymbol{E}^{*}}{2} + \frac{\boldsymbol{U}_i^{n+1} + \boldsymbol{U}_i^{*}}{2}\times\boldsymbol{B}^{n+1}), \\
        &\boldsymbol{U}_e^{n+1} - \boldsymbol{U}_e^{**} = \frac{e\Delta t}{m_er_L}(\frac{\boldsymbol{E}^{n+1} + \boldsymbol{E}^{*}}{2} + \frac{\boldsymbol{U}_e^{n+1} + \boldsymbol{U}_e^{*}}{2}\times\boldsymbol{B}^{n+1}),\\
        &\boldsymbol{E}^{n+1} - \boldsymbol{E}^{*} = -\frac{e\Delta t}{\lambda_D^2r_L}(n_i\frac{\boldsymbol{U}_i^{n+1} + \boldsymbol{U}_i^{*}}{2}- n_e\frac{\boldsymbol{U}_e^{n+1} + \boldsymbol{U}_e^{*}}{2}),\\
        &\boldsymbol{\phi}^{n+1} - \boldsymbol{\phi}^{*} = \frac{\chi\Delta t}{\lambda_D^2r_L}(n_i^{*} - n_e^{*}),
    \end{aligned}
\end{equation*}
which forms a linear system $\boldsymbol{A}\boldsymbol{x}=\boldsymbol{b}$, with
\begin{equation*}
    \boldsymbol{b}=(U_{ix}^{**}, U_{iy}^{**},U_{iy}^{**},U_{ex}^{**}, U_{ey}^{**},U_{ey}^{**},E_x^{*},E_y^{*},E_z^{*})^{T},
\end{equation*}
\begin{equation*}
    \boldsymbol{x}=(U_{ix}^{n+1}, U_{iy}^{n+1},U_{iy}^{n+1},U_{ex}^{n+1}, U_{ey}^{n+1},U_{ey}^{n+1},E_x^{n+1},E_y^{n+1},E_z^{n+1})^{T},
\end{equation*}
and
\begin{equation*}
    \boldsymbol{A}=\left(\begin{array}{ccccccccc}
    1 & -\frac{\alpha B^{n+1}_z}{2} & \frac{\alpha B^{n+1}_y}{2} & 0& 0 & 0 & -\frac{\alpha}{2} & 0 & 0 \\
    \frac{\alpha B^{n+1}_z}{2} & 1 & -\frac{\alpha B^{n+1}_x}{2} & 0& 0 & 0 & 0 & -\frac{\alpha}{2} & 0 \\
    -\frac{\alpha B^{n+1}_y}{2} & \frac{\alpha B^{n+1}_x}{2} & 1 & 0& 0 & 0 & 0 & 0 & -\frac{\alpha}{2} \\
    0 & 0 & 0 & 1 & -\frac{\beta B^{n+1}_z}{2} & \frac{\beta B^{n+1}_y}{2} &  -\frac{\beta}{2} & 0 & 0\\
    0 & 0 & 0 &  \frac{\beta B^{n+1}_z}{2} & 1 &  -\frac{\beta B^{n+1}_x}{2} & 0 &  -\frac{\beta}{2} & 0\\
    0 & 0 & 0 &  -\frac{\beta B_y}{2} &  \frac{\beta B_x}{2} & 1 & 0 & 0 &  -\frac{\beta}{2}\\
    -\frac{\gamma n_i}{2} & 0 & 0 & \frac{\gamma n_e}{2} & 0 & 0 & 1 & 0 & 0\\
    0 & -\frac{\gamma n_i}{2} & 0 & 0 & \frac{\gamma n_e}{2} & 0 & 0 & 1 & 0 \\
    0 & 0 & -\frac{\gamma n_i}{2} & 0 & 0 & \frac{\gamma n_e}{2} & 0 & 0 & 1
    \end{array}\right),
\end{equation*}
where $\alpha = \frac{e\Delta t}{m_ir_L}, \beta = -\frac{e\Delta t}{m_er_L}, \gamma =-\frac{e\Delta t}{\lambda_D^2r_L} $,
$\boldsymbol{B}^{n+1}$ is obtained at the last step. This system can be solved by the Gaussian Elimination method with partial pivoting.

\section{Numerical results}
\label{results}
\subsection{Shock tube}
To understand the characteristic waves of the partially ionized plasma system,
the shock tube with varying plasma ratios is tested. In the neutral limit, no plasma is present in the system, and the model reduces to the classical Sod shock tube problem. In the fully plasma limit, without neutral gas, the model gets to the Brio-Wu shock tube problem of ideal MHD \cite{brio_upwind_1988}.

Within the computational domain is [0.0, 1.0],
the initial condition for the fully ideal MHD equation is
\begin{equation*}
    (\rho, u, p, B_x, B_y)= \begin{cases}(1.0,0,1.0,0.75,1.0) & x \leq 0, \\ (0.125,0,0.1,0.75,-1.0) & x>0.\end{cases}
\end{equation*}
Then, for the partially ionized plasma, the plasma takes a mass fraction of $\theta$, and the neutral gas has a mass fraction of $1-\theta$.
The initial condition for plasma part is changed to
\begin{equation*}
    (\rho , u, p, B_x, B_y)= \begin{cases}(1.0\theta,0,1.0\theta,0.75\sqrt{\theta},1.0\sqrt{\theta}) & x \leq 0, \\ (0.125\theta,0,0.1\theta,0.75\sqrt{\theta},-1.0\sqrt{\theta}) & x>0,\end{cases}
\end{equation*}
and the initial condition for neutral gas is
\begin{equation*}
    (\rho , u, p)= \begin{cases}(1.0(1-\theta),0,1.0(1-\theta)) & x \leq 0, \\ (0.125(1-\theta),0,0.1(1-\theta)) & x>0.\end{cases}
\end{equation*}

\begin{figure}[H]
\centering
\begin{subfigure}[b]{0.32\textwidth}
    \centering
    \includegraphics[width=0.98\textwidth]{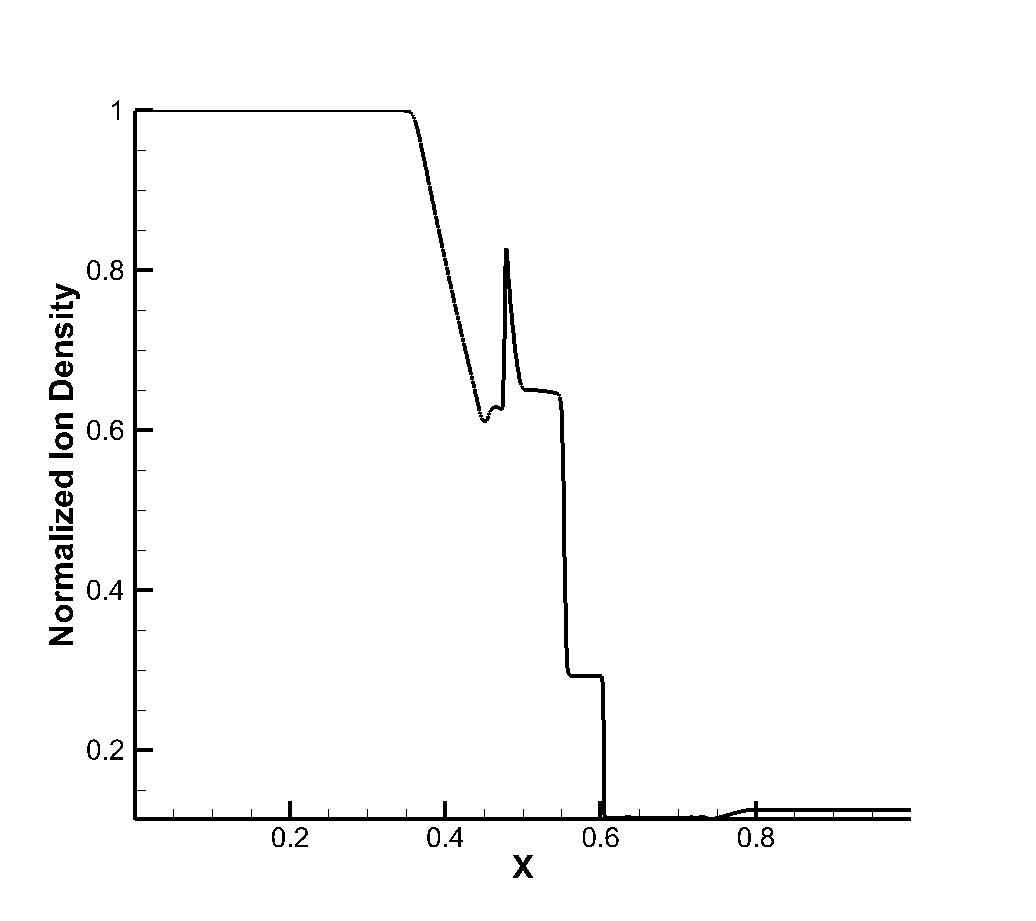}
    \caption{100\% plasma}
    \label{fig:pfig}
\end{subfigure}
\begin{subfigure}[b]{0.32\textwidth}
    \centering
    \includegraphics[width=0.98\textwidth]{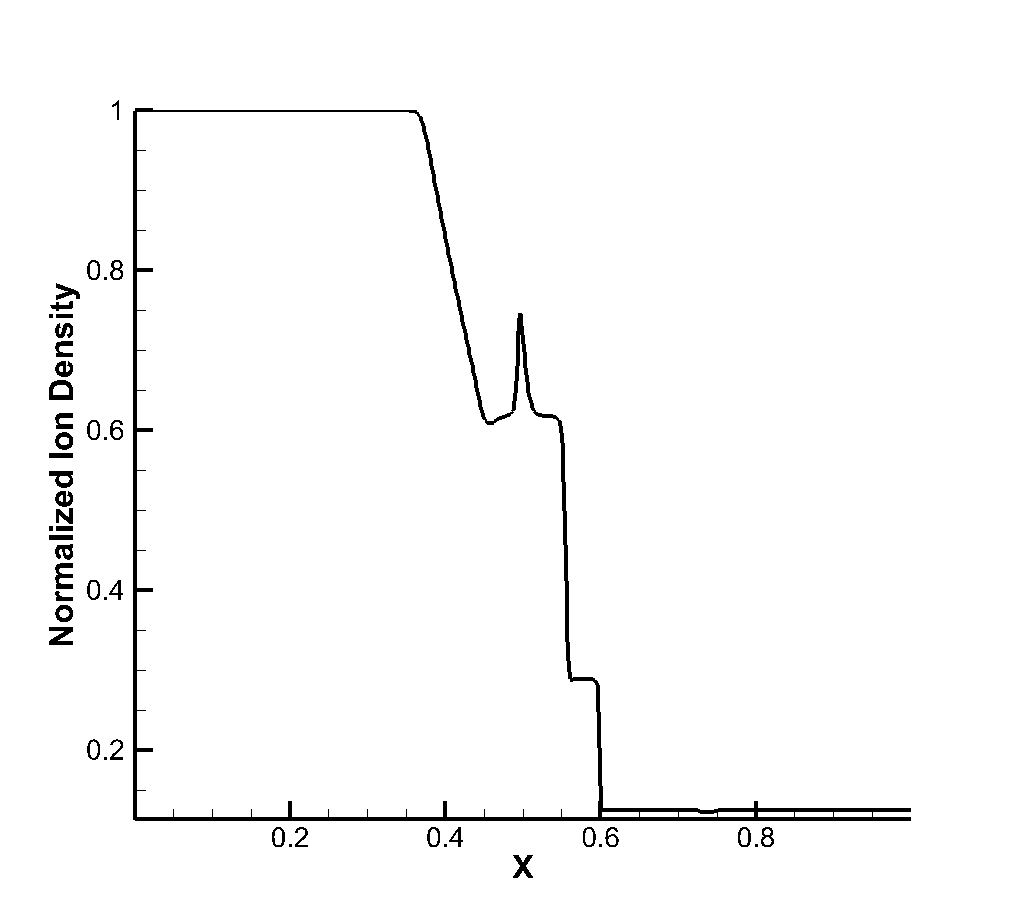}
    \caption{75\% plasma}
    \label{fig:pfilsr}
\end{subfigure}
\begin{subfigure}[b]{0.32\textwidth}
    \centering
    \includegraphics[width=0.98\textwidth]{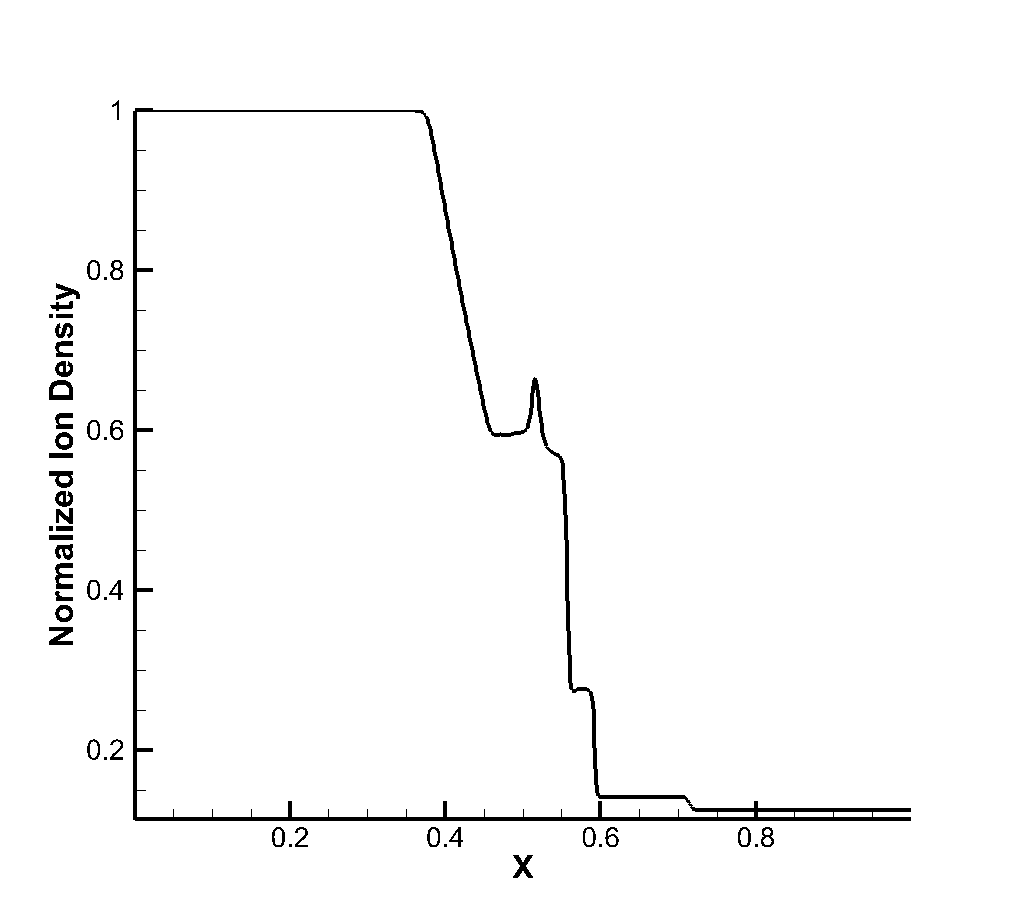}
    \caption{50\% plasma}
    \label{fig:pfilsr}
\end{subfigure}
\vskip\baselineskip
\begin{subfigure}[b]{0.32\textwidth}
    \centering
    \includegraphics[width=0.98\textwidth]{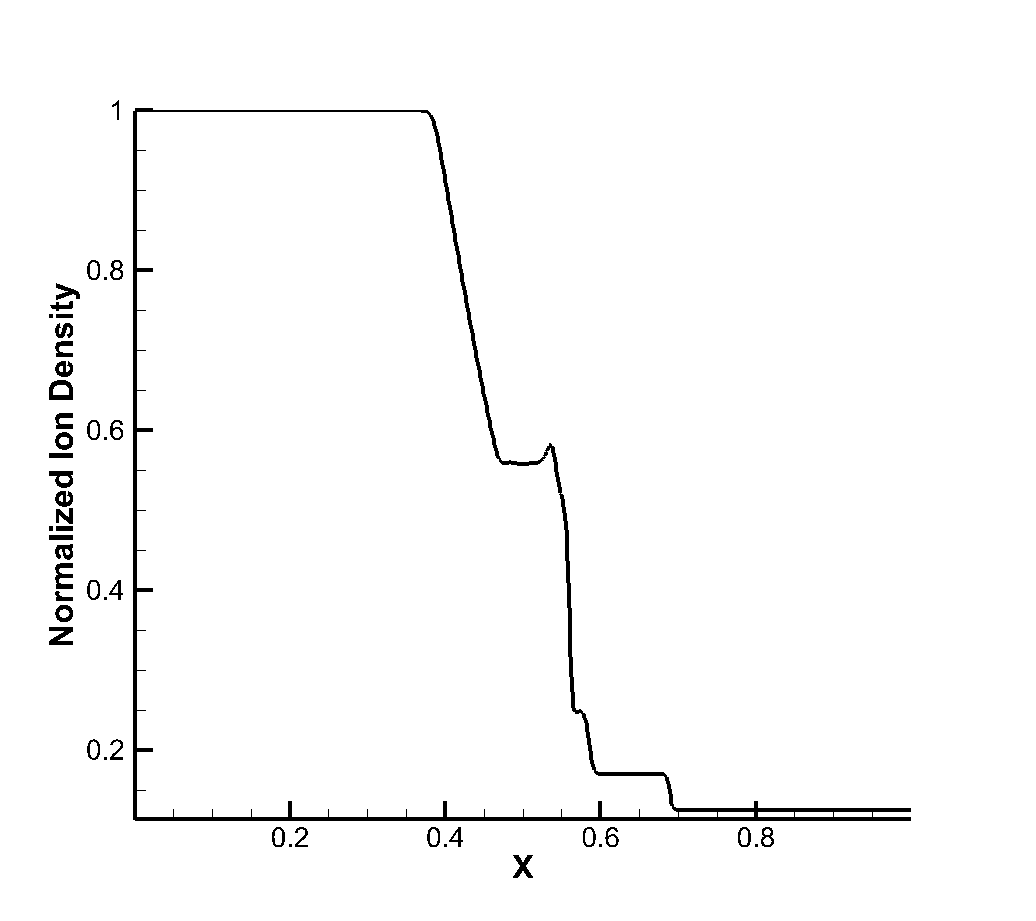}
    \caption{25\% plasma}
    \label{fig:pfirs}
\end{subfigure}
\begin{subfigure}[b]{0.32\textwidth}
    \centering
    \includegraphics[width=0.98\textwidth]{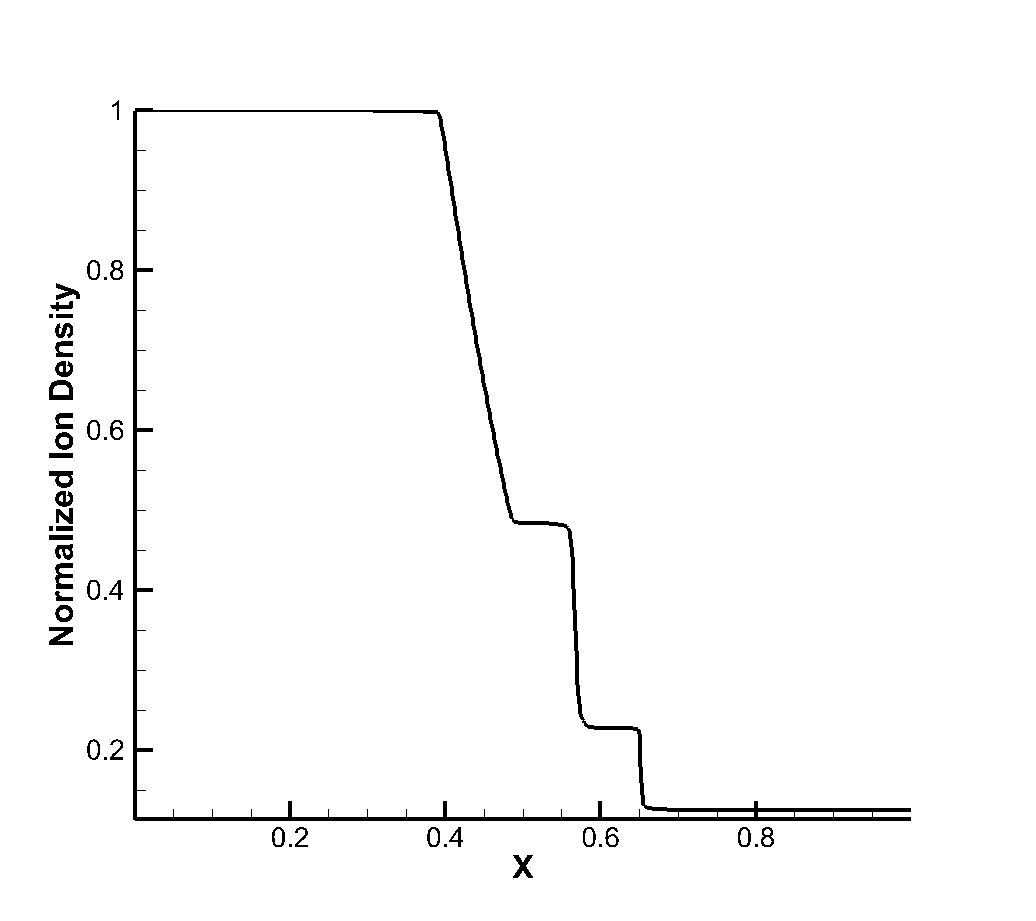}
    \caption{0\% plasma}
    \label{fig:pfirr}
\end{subfigure}
\caption{Normalized density profile of shock tube under different plasma fraction}
\label{fig:briowu}
\end{figure}

\begin{figure}[H]
\centering
\begin{subfigure}[b]{0.32\textwidth}
    \centering
    \includegraphics[width=0.98\textwidth]{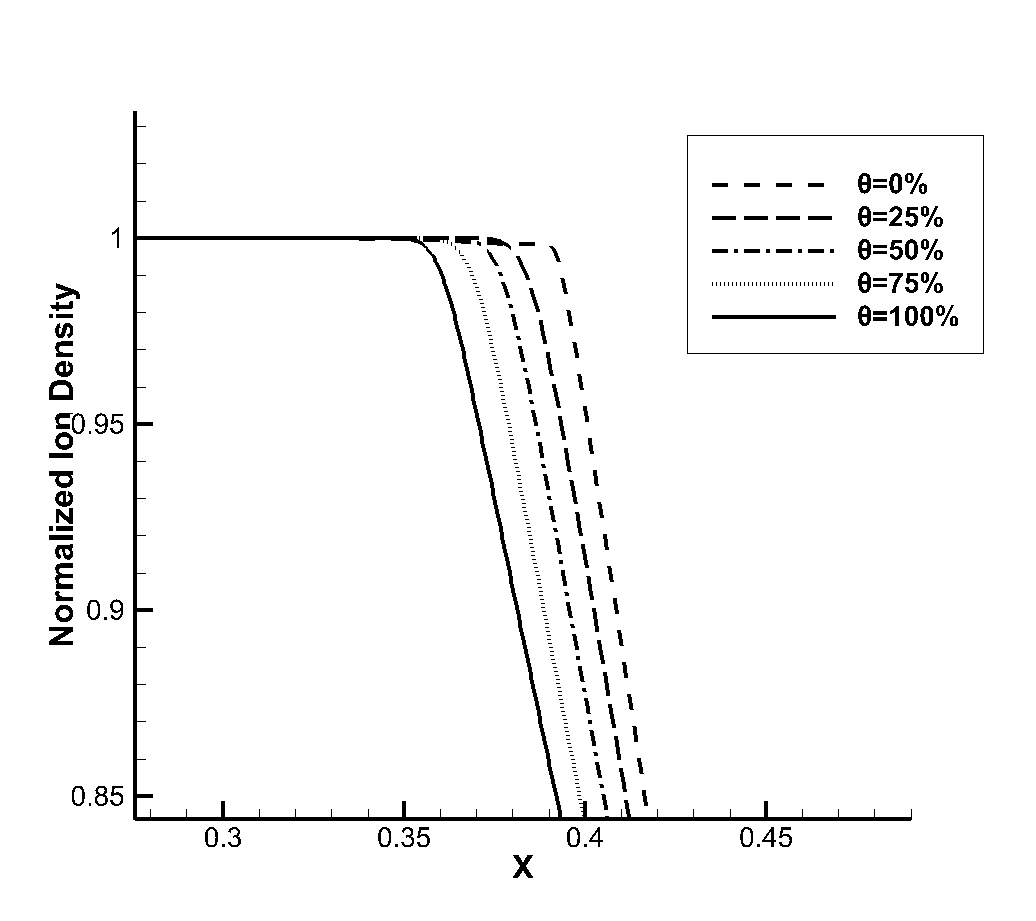}
    \caption{Left-going rarefaction wave }
\end{subfigure}
\begin{subfigure}[b]{0.32\textwidth}
    \centering
    \includegraphics[width=0.98\textwidth]{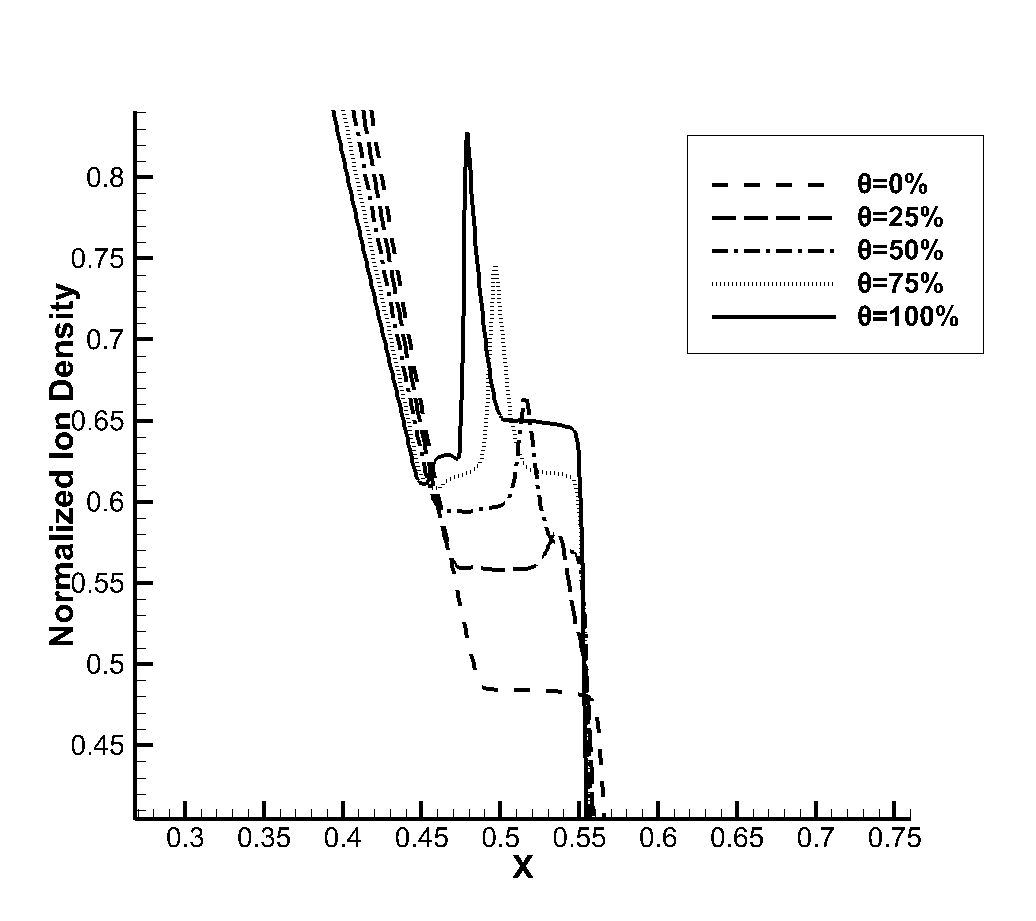}
    \caption{Compound wave}
\end{subfigure}
\begin{subfigure}[b]{0.32\textwidth}
    \centering
    \includegraphics[width=0.98\textwidth]{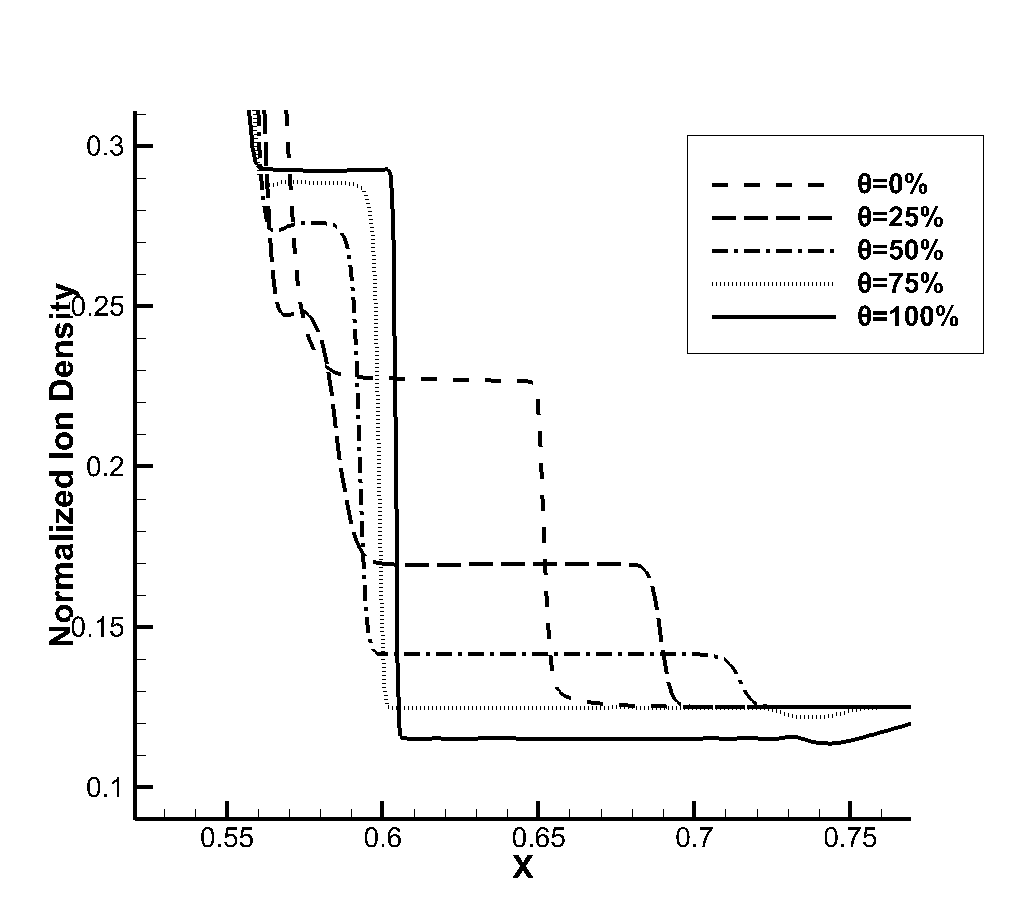}
    \caption{Right-going shock and rarefaction}
\end{subfigure}
\caption{Behaviors of different waves in Fig.\ref{fig:briowu} across different plasma fraction}
\label{fig:briowu_detail}
\end{figure}

Fives cases with distribution of $\theta$, such as 0\%, 25\%, 50\%, 75\%, 100\%, are tested.
As shown in Fig.(\ref{fig:briowu}), as the plasma fraction decreases in the system,
the density profile gradually makes a transition from ideal-MHD solution to the Euler solution.
Specifically, the wave structures change with the following ways:
\begin{enumerate}
    \item The five-wave structure of ideal MHD (two fast magnetosonic rarefaction wave, one slow magnetosonic shock, one contact discontinuity, and one compound wave) gradually shrinks to a three-wave structure of an Euler flow (one acoustic rarefaction, one acoustic shock, and one contact discontinuity).
    \item The compound wave presented in ideal MHD disappears in the Euler limit.
    \item The right-going magnetosonic rarefaction wave present in ideal MHD turns to right-going acoustic shock in the Euler limit.
    \item The left-going magnetosonic rarefaction wave of ideal MHD gradually becomes a left-going acoustic rarefaction wave with a smaller wave speed.
    \item The right-going slow magnetosonic shock of ideal MHD disappears in the Euler limit.
\end{enumerate}
The transitions shown above can be seen more clearly in Fig.(\ref{fig:briowu_detail})

To understand the underlying transition mechanism, characteristic waves' behaviors under different plasma ratios are analyzed. In the Euler limit, $\tau_\alpha\rightarrow0$, electrons, protons, and neutrals are strongly coupled, behaving like a single fluid. Then the governing equation for this single fluid can be written as,
\begin{equation}
\begin{aligned}
&(\rho)_{,t}+(\rho U)_{,x}  =0, \\
&(\rho U)_{,t}+\left(\rho U^2+p_*-B_x^2\right)_{,x} =0, \\
&(\rho V)_{,t}+\left(\rho U V-B_x B_y\right)_{,x} =0, \\
&(\rho W)_{,t}+\left(\rho U W-B_x B_z\right)_{,x} =0, \\
&\left(B_y\right)_t+\left(B_y U-B_x V\right)_{,x} =0, \\
&\left(B_z\right)_t+\left(B_z U-B_x W\right)_{,x} =0, \\
&(\rho \mathscr{E})_{,t}+\left(\left(\rho \mathscr{E}+p_*\right) U-B_x\left(B_x U+B_y V+B_z W\right)\right)_{,x} = 0,
\end{aligned}
\label{1DMHD}
\end{equation}
where, $\rho=\rho_e+\rho_i+\rho_n$ is the total density of electrons, ions, and neutrals. $U = \frac{1}{\rho}(\rho_i U_i+\rho_e U_e + \rho_n U_n)$ is the velocity of the center of mass, so as for velocity $V$ and $W$. $\mathscr{E} = \frac{1}{\rho}(\rho_i \mathscr{E}_i +\rho_e \mathscr{E}_e + \rho_n \mathscr{E}_n)$ is total energy of three species. $p_*=p+\frac{1}{2}\left(B_x^2+B_y^2+B_z^2\right)$ is the total pressure composed of mixture thermal pressure $p=p_i+p_e+p_n$ and magnetic pressure. For an ideal gas in equilibrium, the thermal energy is related to pressure through the relation
$\rho e=p /(\gamma-1)$.
The characteristic speed of seven eigenwaves is,
$$
\lambda_1 = U, \quad \lambda_2 = U+b_x, \quad \lambda_3 = U-b_x,\quad   \lambda_4 = U - c_f, \quad \lambda_5 = U + c_f, \quad \lambda_6 = U - c_s, \quad \lambda_7 = U + c_s
$$
where, $b_x=B_x/\sqrt{\rho}$ is alfven speed. $c_f$ is fast magnetosonic speed,

$$
\begin{aligned}
c_f^2 = \frac{a^2+b^2}{2}(1+\sqrt{\frac{a^4+ b^4+2\gamma p(b_{\perp}^2-{b_x}^2)}{(a^2+b^2)^2}}),\\
c_s^2 = \frac{a^2+b^2}{2}(1-\sqrt{\frac{a^4+ b^4+2\gamma p(b_{\perp}^2-{b_x}^2)}{(a^2+b^2)^2}}),
\end{aligned}
$$
where $b^2=b_x^2+b_y^2+b_z^2$, $b_{\perp}^2=b_y^2+b_z^2$, $a=\sqrt{\gamma p/\rho}$ is sound speed of mixture. In this 1D case, velocity $V$ and $W$ are passively transported, so it doesn't appear in the eigensystem.

The impacts of plasma fraction $\theta$ on the behaviors of the system are analyzed below. Given a plasma fraction $\theta$, Alfven speed $b_x =\sqrt{\theta}(B_x/\sqrt{\rho})$. In neutral limit where $\theta=0$, $b_x = b_y = b_z = 0$, fast magnetosonic speed $c_f = a$, and slow magnetosonic speed $c_s = 0$. This explains the phenomena shown in Fig.(\ref{fig:briowu_detail}), where the left-going fast magnetosonic rarefaction wave gradually turns to an acoustic rarefaction wave, the right-going fast shock becomes acoustic shock, and the slow shock goes to contact discontinuity and disappears. In plasma limit where $\theta=1$, the characteristic system completely becomes that of ideal-MHD.

\subsection{Orszag-Tang vortex}
In this section, the Orszag-Tang Vortex problem was tested to explore how neutrals influence MHD shocks. This problem was originally designed to study the MHD turbulence\cite{orszag1979small}. It was intensively studied later and gradually becomes a benchmark problem of 2D MHD codes to test the capability to handle the formation of MHD shocks and shock-shock interactions\cite{londrillo_highorder_2000,dahlburg_evolution_1989,zachary_higher-order_1994,jiang_high-order_1999}. The computational domain is $[0,2\pi]\times[0,2\pi]$, and the initial conditions are:
\begin{center}

\begin{tabular}{ |p{3cm}||p{3cm}|p{3cm}|p{3cm}|  }
 \hline
 \multicolumn{4}{|c|}{Initial condition} \\
 \hline
 Item & Ions & Electrons & Neutrals\\
 \hline
 m& 1.0 & 0.04 & 1.0 \\
 n& $\gamma^2\theta $ & $\gamma^2\theta $  & $\gamma^2(1-\theta) $\\
 p & $\gamma(\theta/(1+\theta)) $& $\gamma(\theta/(1+\theta))$&  $\gamma(1-\theta)/(1+\theta)$\\
 $V_x$ &$-\sin(y) $& $-\sin(y)$&  $-\sin(y)$\\
 $V_y$ &$\sin(x) $ & $\sin(x)$&$\sin(x)$\\
 $B_y$ & $\sin(2x)\sqrt{\theta} $ & $\sin(2x)\sqrt{\theta}$   & $-$\\
 \hline
\end{tabular},
\end{center}
where $\gamma=5/3$ is specific heat capacity and $\theta$ is fraction of ions. Three cases with $\theta=1,0.75,0$ are tested.

\begin{figure}[H]
\centering
\begin{subfigure}[b]{0.48\textwidth}
\centering
    \centering
    \includegraphics[width=0.98\textwidth]{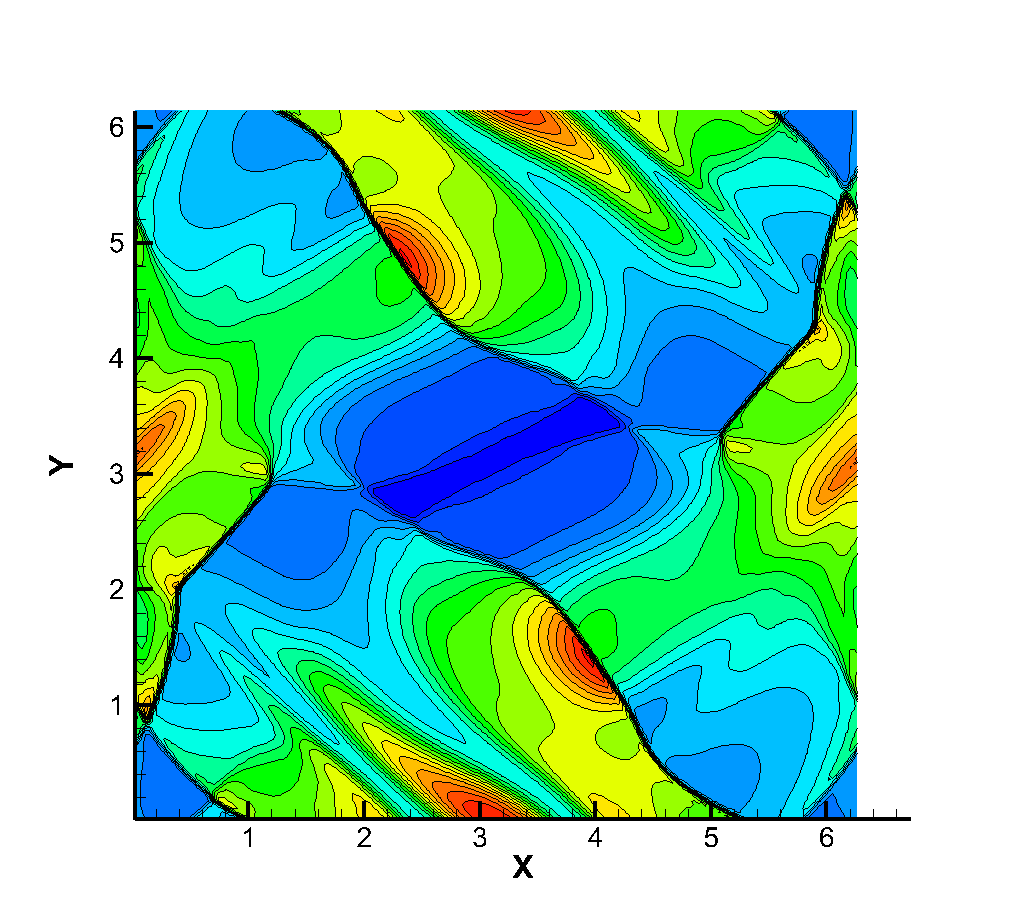}
\caption{100\% plasma t = 2}
\label{fig:orszagpip100t2}
\end{subfigure}
\begin{subfigure}[b]{0.48\textwidth}
\centering
    \centering
    \includegraphics[width=0.98\textwidth]{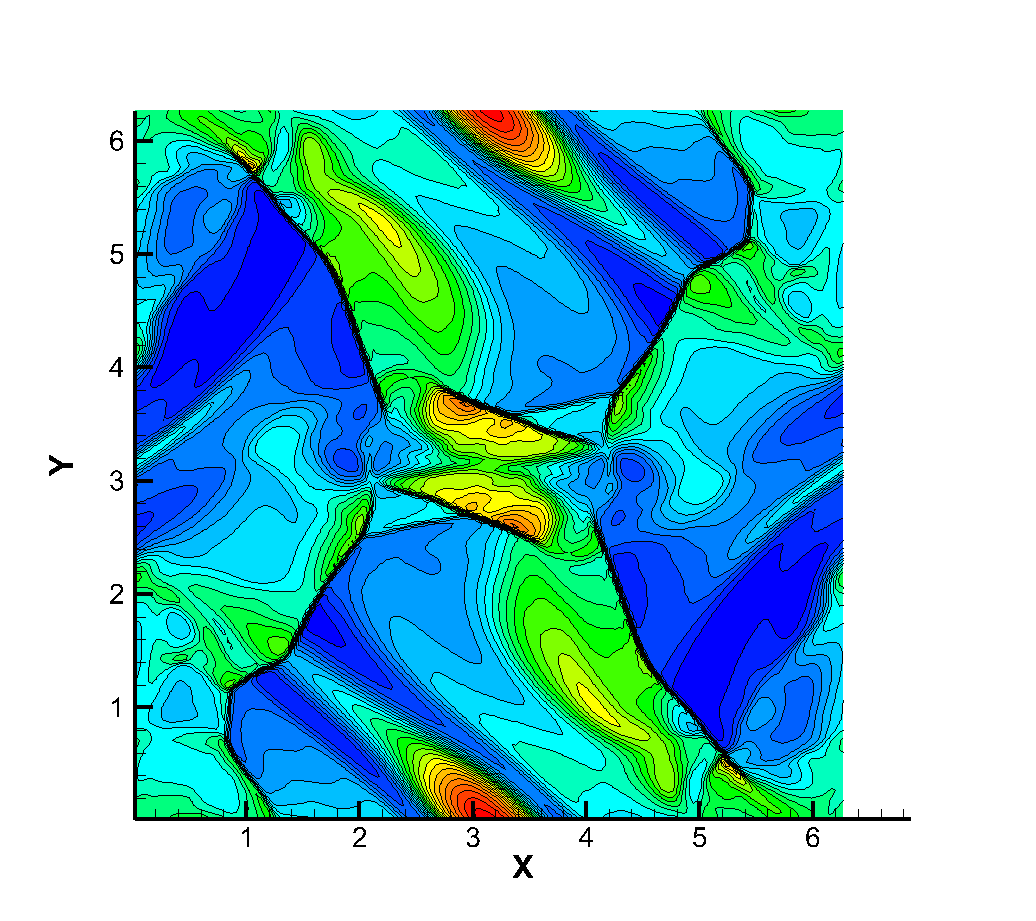}
\caption{100\% plasma t = 3}
\label{fig:orszagpip100t3}
\end{subfigure}
\caption{Orszag-Tang vortex in partially ionized plasma with full ionization}
\label{fig:orszagpip100}
\end{figure}

\begin{figure}[H]
\centering
\begin{subfigure}[b]{0.48\textwidth}
\centering
    \centering
    \includegraphics[width=0.98\textwidth]{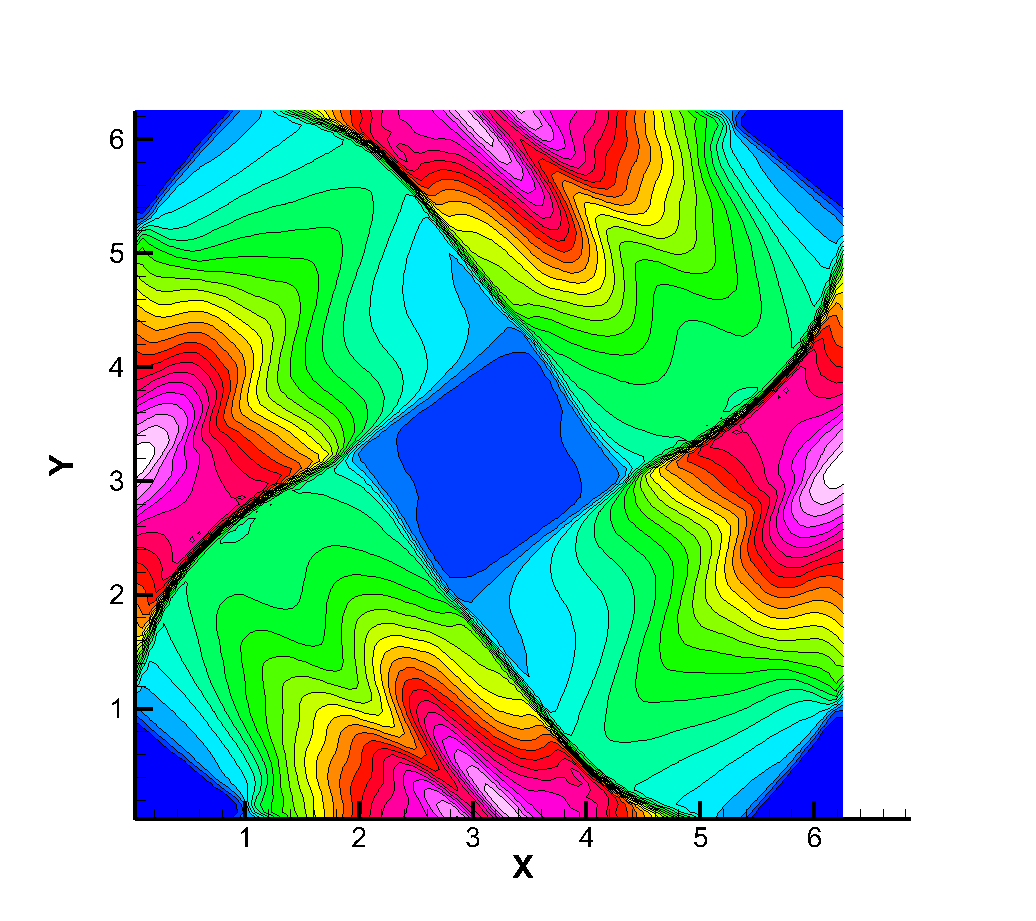}
\caption{75\% plasma t = 2}
\label{fig:orszagpip75t2}
\end{subfigure}
\begin{subfigure}[b]{0.48\textwidth}
\centering
    \centering
    \includegraphics[width=0.98\textwidth]{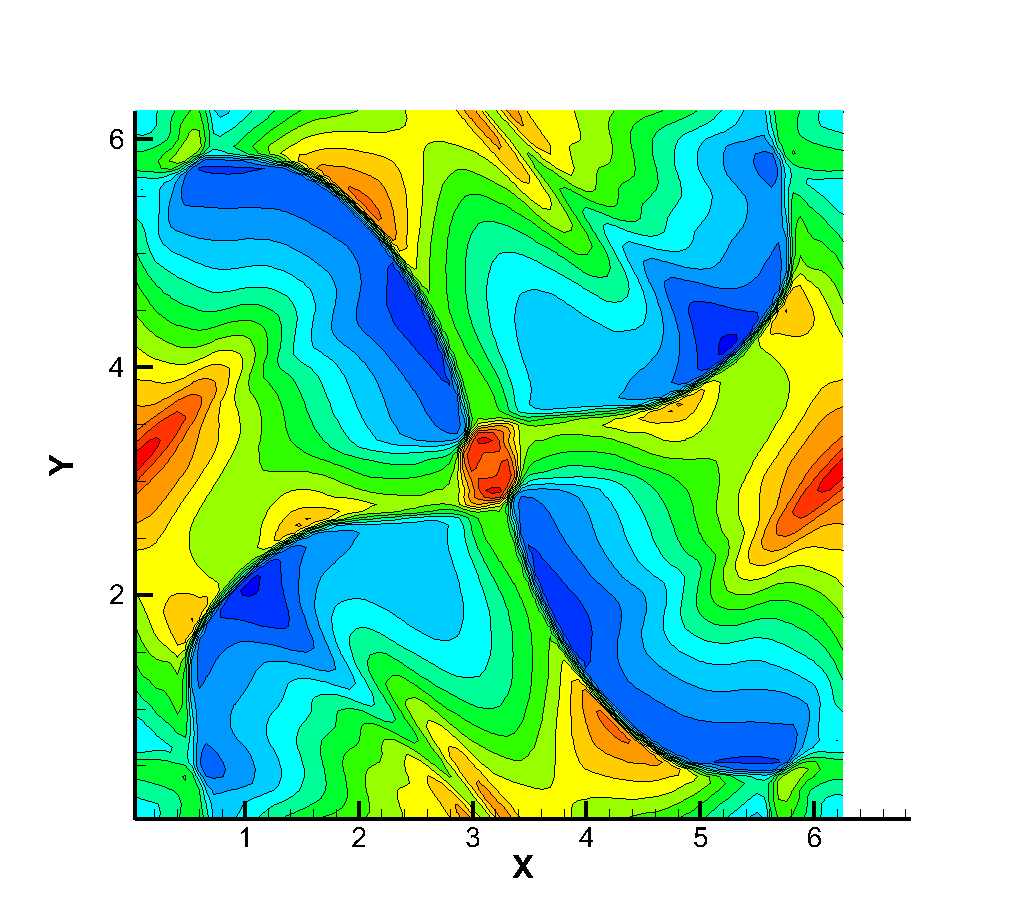}
\caption{75\% plasma t = 3}
\label{fig:orszagpip75t3}
\end{subfigure}
\caption{Orszag-Tang vortex in partially ionized plasma with 75\% ionization}
\label{fig:orszagpip75}
\end{figure}

\begin{figure}[H]
\centering
\begin{subfigure}[b]{0.48\textwidth}
\centering
    \centering
    \includegraphics[width=0.98\textwidth]{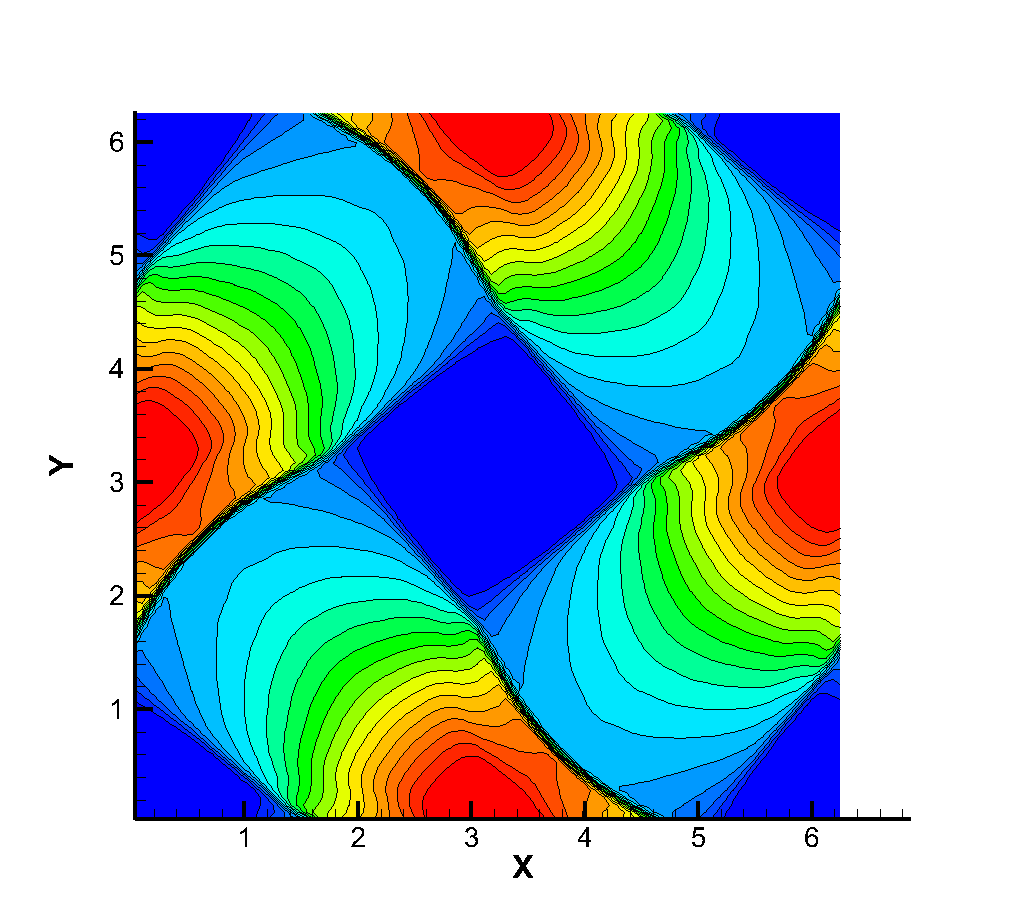}
\caption{0\% plasma t = 2}
\label{fig:orszagpip0t2}
\end{subfigure}
\begin{subfigure}[b]{0.48\textwidth}
\centering
    \centering
    \includegraphics[width=0.98\textwidth]{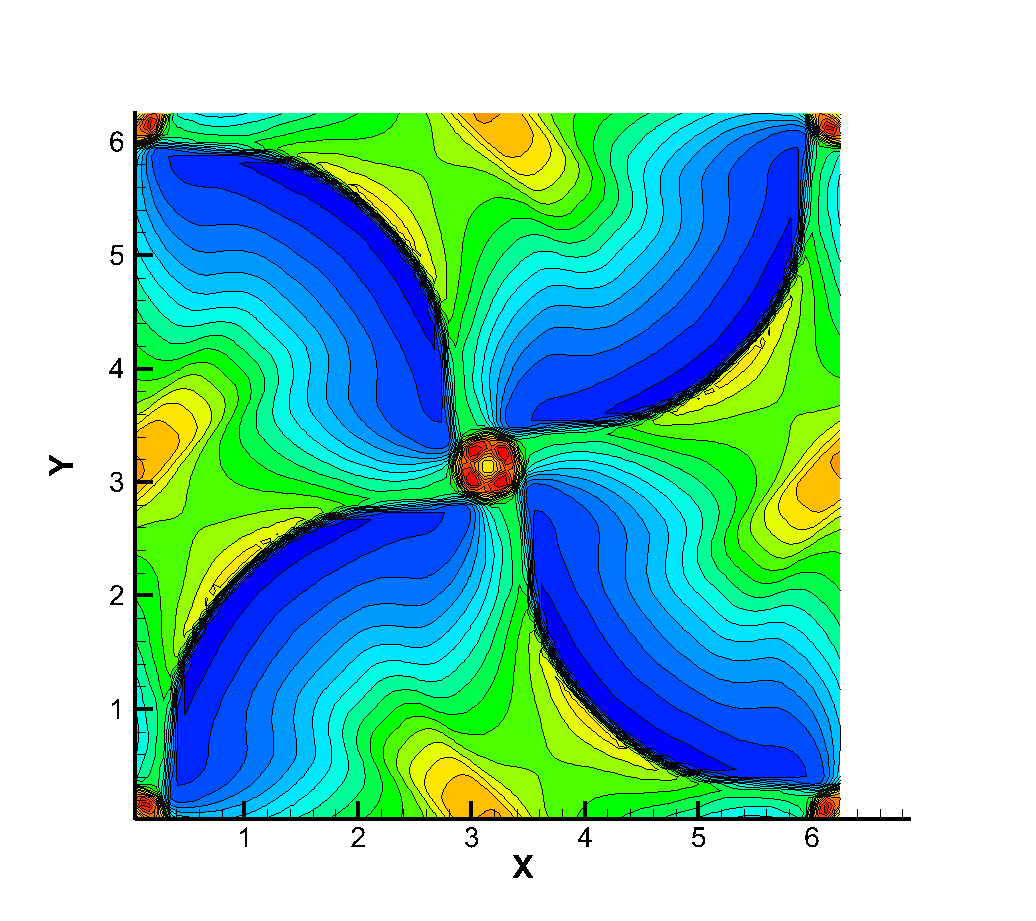}
\caption{0\% plasma t = 3}
\label{fig:orszagpip0t3}
\end{subfigure}
\caption{Orszag-Tang vortex in partially ionized plasma with 0\% ionization}
\label{fig:orszagpip0}
\end{figure}

The flow structures from the fully ionized plasma to a fully neutral gas case are shown in Fig.\ref{fig:orszagpip100}-Fig.\ref{fig:orszagpip0}. Comparing Fig.\ref{fig:orszagpip100t2}, Fig.\ref{fig:orszagpip75t2} and Fig.\ref{fig:orszagpip0t2}, it shows that the compound shock \cite{jiang_high-order_1999} at the location from $(x,y) \sim (3\pi/2,0) \text{to} (\pi,3\pi/4)$ gradually disappear and transforms to normal acoustic shock. Comparing Fig.\ref{fig:orszagpip100t3}, Fig.\ref{fig:orszagpip75t3} and Fig.\ref{fig:orszagpip0t3}, it is found that the slow shock front at the location from $(x,y) \sim (\pi,3\pi/4) \text{to} (\pi/2,\pi)$ gradually disappear. The fast shock located at $(x,y) \sim (3\pi/2,\pi/2) \text{to} (5\pi/4,3\pi/4)$ turns to normal acoustic shock front. Therefore, similar phenomena are observed as the shock tube case in this more complicated 2D simulation.


\section{Conclusions}
\label{conclusions}

In summary, this work presents kinetic models for multi-species transport and interaction among neutrals, electron, and proton, along with the electromagnetic field in the coupled evolution of the Euler and ideal MHD equations in the continuum flow regime.
The 1D Riemann problem is solved for the partial ionized plasma to explore nonlinear wave behaviors with the variation of the mass fraction of the plasma. As a result, the Euler solutions and the ideal MHD solutions are obtained in the limiting cases.
It is observed that in the transition from the MHD to the Euler solutions as a function of $\theta$, the fast magneto-sonic wave transits to the sound wave, and slow magneto-sonic wave and compound wave disappear.
The Orszag-Tang vortex test case is also used to study the influence of neutral gas on MHD wave structure and similar trend as the 1D case
has been confirmed.  Based on the current kinetic formulation, the multiscale method for the non-equilibrium PIP system will be constructed in the future.


\section*{Acknowledgements}

This work was supported by National Key R$\&$D Program of China (Grant Nos. 2022YFA1004500), National Natural Science Foundation of China (12172316),
and Hong Kong research grant council (16208021,16301222).

\bibliographystyle{elsarticle-num}
\bibliography{els-v0.7}


\end{document}